\documentclass[a4paper,11pt]{article}

\usepackage{amsmath,amssymb,mathtools}
\usepackage{color}
\usepackage{graphicx}
\usepackage{subfigure}
\usepackage{cite}
\usepackage[colorlinks=true,linkcolor=red,citecolor=blue,urlcolor=blue,bookmarks]{hyperref}
\usepackage{multirow,makecell}
\usepackage{textcomp}
\usepackage{wasysym}
\usepackage{ulem}

\usepackage{verbatim}

\usepackage[utf8]{inputenc}
\usepackage[T1]{fontenc}

\usepackage[text={17.1cm,24.6cm},centering]{geometry} 

\numberwithin{equation}{section}

\def \be {\begin{equation}}
\def \ee {\end{equation}}
\def \ba {\begin{array}}
\def \ea {\end{array}}
\def \bea {\begin{eqnarray}}
\def \eea {\end{eqnarray}}

\def \g {\gamma}

\def \G {\Gamma}
\def \d {\delta}

\def \ve {\varepsilon}
\def \ce {\varepsilon}

\def \s {\sigma}

\def \r {\rho}

\def \cH {\mathcal H}

\def \f {\frac}

\def \lt {\left}
\def \rt {\right}

\def \sr {\sqrt}
\def \td {\tilde}

\def \inf {\infty}

\def \lag {\langle}
\def \rag {\rangle}

\def \ii {\mathrm{i}}

\def \arctanh {\mathop{\rm arctanh}}

\def \tr {\textrm{tr}}

\def \and {{~\textrm{and}~}}

\def \NS {{\textrm{NS}}}
\def \R {{\textrm{R}}}

\def \NS {{\textrm{NS}}}
\def \R {{\textrm{R}}}

\begin{document}

\title{
\textbf{Efficient computation of average subsystem Bures distance between fermionic Gaussian states}
}


\author{
Zhouhao Guo${}^1$,
M.~A.~Rajabpour${}^2$,
and
Jiaju Zhang${}^1$\footnote{Corresponding author: jiajuzhang@tju.edu.cn}
}
\date{}
\maketitle
\vspace{-10mm}
\begin{center}
{\it
${}^1$Center for Joint Quantum Studies and Department of Physics, School of Science, Tianjin University,\\
135 Yaguan Road, Tianjin 300350, China\\
${}^2$Instituto de F\'isica, Universidade Federal Fluminense,\\
Av.~Gal.~Milton Tavares de Souza s/n, Gragoat\'a, 24210-346, Niter\'oi, RJ, Brazil
}
\vspace{10mm}
\end{center}


\begin{abstract}

  The average subsystem trace distance has been proposed as an indicator of quantum many-body chaos and integrability. However, evaluating it presents two main difficulties: high computational cost for large systems and ambiguities in defining and ordering eigenstates in integrable systems. In this work, we develop an efficient algorithm to compute the Bures distance between fermionic Gaussian states, enabling access to larger system sizes. Using this method, we calculate the average subsystem Bures distance for eigenstates in the spin-1/2 transverse-field Ising chain and the Dirac fermion formulation of the quadratic Sachdev-Ye-Kitaev (Dirac SYK$_2$) model, as well as for random pure fermionic Gaussian states. To handle degeneracy in the Ising chain, we consider simultaneous eigenstates of all local conserved charges and employ these charges to systematically order degenerate states. Our results are consistent with the earlier conjecture of a linear growth with subsystem size. We show that the distinct scaling of the average subsystem distances in chaotic versus integrable systems originates from discontinuities of local conserved charges across the spectrum in integrable models. For the Dirac SYK$_2$ model and random pure Gaussian states, we obtain similar results for the average subsystem distances, which do not show a linear increase.

\end{abstract}

\baselineskip 18pt
\thispagestyle{empty}
\newpage


\tableofcontents

\section{Introduction}\label{sectionIntroduction}

The precise definition of quantum many-body chaos and integrability has been a long-standing problem \cite{Caux:2010by}.
A widely used indicator is level spacing statistics \cite{Berry:1977kiw,Bohigas:1984vrm}, which, however, can lead to ambiguous or even incorrect classifications \cite{Finkel:2005yof,Sieberer:2019xfn}.
The out-of-time-ordered correlator has also been proposed as an indicator of chaos \cite{Larkin:1969ifp,Maldacena:2015waa}, but counterexamples still exist \cite{Kukuljan:2017xag,Rozenbaum:2019nwn}.
Another approach uses the scaling behavior of the average entanglement entropy in the spectrum to distinguish integrable and chaotic systems \cite{Vidmar:2017uux,Vidmar:2017pak,Vidmar:2018rqk,Hackl:2018tyl,Jafarizadeh:2019xxc,LeBlond:2019eoe,Kliczkowski:2023qmp},
though it often requires relatively large system sizes to clearly observe the difference.
The average entanglement entropy in chaotic models \cite{Vidmar:2017pak,Kliczkowski:2023qmp} is universal and consistent with the Page curve \cite{Page:1993df,Foong:1994eja,Sanchez-Ruiz:1995bhf,Sen:1996ph} for a random state in the Hilbert space. In integrable models \cite{Vidmar:2017uux,Vidmar:2018rqk,Hackl:2018tyl,Jafarizadeh:2019xxc,LeBlond:2019eoe}, it also follows a volume law but is apparently smaller. The average entanglement entropy in quantum chaotic quadratic Hamiltonians \cite{Liu:2017kfa,Zhang:2020kia,Lydzba:2020qfx,Lydzba:2021hml}, i.e. models exhibiting single-particle quantum chaos, differs from that in integrable models and matches the universal result for random pure fermionic Gaussian states \cite{Bianchi:2021lnp,Bianchi:2021aui}.

Recently, the average subsystem trace distance was proposed as an indicator of quantum many-body chaos and integrability \cite{Khasseh:2023kxw}.
For two quantum states with density matrices $\r$ and $\s$, the trace distance is defined as \cite{Nielsen:2010oan}
\be
D(\r,\s)=\f12 \tr|\r-\s|.
\ee
By definition, $0\leq D(\r,\s)\leq 1$.
We focus on a one-dimensional system of total size $L$ containing a contiguous $A$ with size $\ell$.
We denote the ordered eigenstates as $|i\rangle$, $i = 1, 2, \cdots, d$, in the $d$-dimensioanl full Hilbert space or within a symmetry sector of dimension $d$.
For two consecutive eigenstates $|i\rag$ and $|i+1\rag$, one can choose a continuous subsystem $A$ of size $\ell$ and calculate the subsystem trace distance $D(\r_{A,i},\r_{A,i+1})$ between their reduced density matrices $\r_{A,i}$ and $\r_{A,i+1}$ and then define the average subsystem trace distance
\be
\lag D_A \rag = \f{1}{d-1} \sum_{i=1}^{d-1} D(\r_{A,i},\r_{A,i+1}).
\ee

In chaotic systems, it is conjectured that the average subsystem trace distance vanishes in the scaling limit $L\to+\inf$ with fixed ratio $x=\f{\ell}{L}<\f12$, which is consistent with eigenstate thermalization hypothesis (ETH) \cite{Deutsch:1991msp,Srednicki:1994mfb,Rigol:2007mja,Dymarsky:2016ntg} and is the same as the average subsystem trace distance between random states \cite{Kudler-Flam:2021rpr,Kudler-Flam:2021alo,deMiranda:2022rze}.
In integrable systems, the average subsystem trace distance remains finite for fixed $x$.
The examples in \cite{Khasseh:2023kxw} indicate that the average subsystem trace distance takes a universal linear increase for $x<\f12$ in a large class of integrable systems. However, it is difficult to draw a definitive conclusion for two reasons.
First, computing the subsystem trace distance is numerically demanding, and existing results are limited to small sizes around $L = 20$, $\ell = 12$.
Second, a large number of energy degeneracies leads to ambiguities in defining and ordering the eigenstates.

In this paper, we address the above two problems in the transverse field Ising chain.
Firstly, we replace the trace distance with the Bures distance and develop an efficient numerical algorithm to calculate the Bures distance between two Gaussian states, allowing us to study much larger system sizes, up to $L=29$, $\ell=28$.
The Bures distance between two state $\r$ and $\s$ is given by \cite{Bures:1969rqp}%
\footnote{In the literature, people also define the fidelity in a little different way
\[
\td F(\rho,\sigma)= \Big( \tr \sqrt{\sqrt{\rho}\sigma \sqrt{\rho}} \Big)^2,
\]
but the defined Bures distance is always the same
\[
B(\r,\s) = \sqrt{2\Big(1-\sqrt{\td F(\r,\s)}\Big)}.
\]%
}
\be
B(\r,\s) = \sqrt{2(1-F(\r,\s))},
\ee
with the fidelity being defined as \cite{Jozsa:1994mnd,Nielsen:2010oan}
\begin{equation}
    F(\rho,\sigma)=\operatorname{tr}\sqrt{\sqrt{\rho}\sigma \sqrt{\rho}}.
    \label{fidelity}
\end{equation}
By definition $0\leq B(\r,\s)\leq \sqrt{2}$. For the whole spectrum of some symmetry sector, we define the average subsystem Bures distance
\be
\lag B_A \rag = \f{1}{d-1} \sum_{i=1}^{d-1} B(\r_{A,i},\r_{A,i+1}).
\ee
The trace distance and fidelity between two states \(\r\) and \(\s\) are linked via the Fuchs-van~de~Graaf inequalities \cite{Nielsen:2010oan}
\be
1 - F(\r, \s) \leq D(\r, \s) \leq \sqrt{1 - F(\r, \s)^2},
\ee
from which we establish bounds for the Bures distance in terms of the trace distance
\be
\sqrt{2\big(1 - \sqrt{1 - D(\r, \s)^2}\big)} \leq B(\r, \s) \leq \sqrt{2D(\r, \s)}.
\ee
By averaging and utilizing the concavity and convexity of relevant functions, we derive bounds for the average subsystem Bures distance based on the average subsystem trace distance
\be \label{bounds}
\sqrt{2\big(1 - \sqrt{1 - \lag D_A \rag^2}\big)} \leq \lag B_A \rag \leq \sqrt{2\lag D_A \rag}.
\ee
These inequalities provide a consistency check for the numerical results.

Secondly, we use simultaneous eigenstates of all the local conserved charges and resolve the degeneracy by ordering the states according to their eigenvalues.
Our results seem consistent with the conjectured linear growth with the subsystem size proposed in \cite{Khasseh:2023kxw}.
Additionally, we show that the differing scaling behaviors of average subsystem trace and Bures distances in chaotic and integrable systems stem from discontinuities of local conserved charges in the spectrum of the integrable systems.

We compute the average subsystem distances in the Dirac fermion formulation of the quadratic Sachdev-Ye-Kitaev model (Dirac SYK$_2$), a particle-number conserving quantum-chaotic quadratic system, and also for random pure fermionic Gaussian states. In both cases, we observe no universal linear growth. The results for the Dirac SYK$_2$ model differ from that in integrable models, but match the universal behavior for random pure Gaussian states. Clarifying the precise relationships among average subsystem distances in integrable models, quantum chaotic quadratic models, and random pure Gaussian states requires further investigation.

The remaining part of the paper is arranged as follows.
In section~\ref{sectionBures} we present an efficient algorithm of calculating the Bures distance between two Gaussian states.
In section~\ref{sectionIsing} we show how to use the local charges to lift the degeneracy of the eigenstates and present the numerical results for the average subsystem trace and Bures distances in transverse field Ising chain.
In section~\ref{sectionSYK2}, we evaluate the average subsystem trace and Bures distances in the Dirac SYK$_2$ model.
In section~\ref{sectionRandom}, we calculate the average subsystem trace and Bures distances between random pure fermionic Gaussian states.
We conclude with discussions in section~\ref{sectionDiscussion}.
In appendix~\ref{appSYK2}, we calculate the average subsystem distance in the sector with fixed particle number \(N=1\) of the Dirac SYK\(_2\) model.
In appendix~\ref{appXXZ}, we consider the average subsystem distances in spin-1/2 XXZ chain.
In appendix~\ref{appSlo}, we discuss the universality of the slope 2.

\section{Bures distance between two Gaussian states} \label{sectionBures}

In this section, we show how to efficiently compute the Bures distance between two Gaussian states using the correlation matrix.
Since the Bures distance is defined in terms of fidelity, we focus on computing fidelity in the following.

For a system described by Pauli matrices, one can apply the Jordan-Wigner transformation to obtain $2\ell$ Majorana fermion modes from $\ell$ sets of Pauli matrices
\begin{equation}
d_{2j-1} = \Big( \prod_{i=1}^{j-1} \sigma^{z}_{i} \Big) \sigma^{x}_{j}, ~~
d_{2j} = \Big( \prod_{i=1}^{j-1} \sigma^{z}_{i} \Big) \sigma^{y}_{j}, ~~
1 \leq j \leq \ell.
\end{equation}
A general Gaussian state with density matrix $\r_\G$ is uniquely characterized by the $2\ell \times 2\ell$ correlation matrix
\be
\G_{m_1m_2} = \tr( \r_\G d_{m_1} d_{m_2} ) - \d_{m_1m_2}, ~~ m_1,m_2=1,2,\cdots,2\ell,
\ee
which is antisymmetric and purely imaginary. The density matrix $\r_\G$ takes the form \cite{Fagotti:2010yr}
\begin{align}
\rho_\G=\frac{1}{Z_\G}\exp\Big(-\frac{1}{2}\sum_{m_1,m_2=1}^{2\ell}W_{m_1m_2}d_{m_1}d_{m_2}\Big),
\end{align}
where
\be
Z_\G=\sqrt{\det\frac{2}{1 + \Gamma}}, ~~
W=\arctanh\G.
\ee
A useful formula in \cite{Fagotti:2010yr} is that
\be \label{rG1rG2}
\r_{\G_1}\r_{\G_2}=\tr(\r_{\G_1}\r_{\G_2})\r_{\G_1\times\G_2},
\ee
with
\be \label{trrG1rG2}
\tr(\r_{\G_1}\r_{\G_2})=\sqrt{\det\f{1+\G_1\G_2}{2}},
\ee
\be \label{G1timesG2}
\G_1\times\G_2 = 1 - (1-\G_1)\f{1}{1+\G_2\G_1}(1-\G_2).
\ee

For general cases with large $\ell$, instead of using directly the $2^\ell\times2^\ell$ density matrices $\r_\G$ to calculate the fidelity, which is a formidable task, we employ the $2\ell\times2\ell$ correlation matrices $\G$, which allow us to access significantly larger system sizes.
Based on the definition of fidelity (\ref{fidelity}) and the properties of Gaussian states, one can exactly compute the fidelity as given in \cite{Banchi:2013uht}
\be \label{FrG1rG2I}
F(\r_{\G_1},\r_{\G_2}) =
\Big(\det\f{1+\G_1}{2}\Big)^{1/4}
\Big(\det\f{1+\G_2}{2}\Big)^{1/4}
\Big[\det\Big(1+\sr{\sr{\f{1-\G_1}{1+\G_1}}\f{1-\G_2}{1+\G_2}\sr{\f{1-\G_1}{1+\G_1}}}\Big)\Big]^{1/2}.
\ee
Recently, an equivalent but more efficient formula for the fidelity was proposed in \cite{Baldwin:2022cjb}
\begin{equation}
  \label{Frhosigma}
F(\rho,\sigma)=\operatorname{tr}\sqrt{\rho\sigma},
\end{equation}
which enables a simplified computation of the fidelity using the correlation matrices
\be \label{FrG1rG2II}
F(\r_{\G_1},\r_{\G_2}) =
\Big(\det\f{1+\G_1}{2}\Big)^{1/4}
\Big(\det\f{1+\G_2}{2}\Big)^{1/4}
\Big[\det\Big(1+\sr{\f{1-\G_1}{1+\G_1}\f{1-\G_2}{1+\G_2}}\Big)\Big]^{1/2}.
\ee
When neither $\G_1$ nor $\G_2$ has eigenvalues that are exactly or very close to $\pm1$, we can directly evaluate the fidelity using (\ref{FrG1rG2II}).
Otherwise, a numerical issue involving the indeterminate form $0 \times \infty$ arises.
To avoid this issue, one can introduce a small cutoff, as done in \cite{Zhang:2022nuh}, which introduces additional numerical errors and requires high-precision arithmetic.
In the remainder of this section, we present an alternative and more efficient method to resolve this issue.

We first consider two special cases. When the dimension of the Hilbert space is two (i.e., $\ell = 1$), the correlation matrices and density matrices take the following forms, respectively \cite{Vidal:2002rm,Latorre:2003kg}
\be
\G_1 = \bigg( \ba{cc} & \ii \g_1 \\ -\ii \g_1 & \ea \bigg), ~~
\G_2 = \bigg( \ba{cc} & \ii \g_2 \\ -\ii \g_2 & \ea \bigg),
\ee
\be
\r_{\G_1} = \f12 \bigg( \ba{cc} 1+\g_1 & \\ & 1-\g_1 \ea \bigg), ~~
\r_{\G_2} = \f12 \bigg( \ba{cc} 1+\g_2 & \\ & 1-\g_2 \ea \bigg).
\ee
The parameters $\gamma_1$, $\gamma_2$ are related to the correlation functions of the two Majorana fermions $d_1$, $d_2$ as $\langle d_1 d_2 \rangle_{\rho_{\Gamma_1}} = \ii\gamma_1$ and $\langle d_1 d_2 \rangle_{\rho_{\Gamma_2}} = \ii\gamma_2$.
Using (\ref{Frhosigma}), we obtain the fidelity
\be \label{Fdeq2}
F(\r_{\G_1},\r_{\G_2}) = \f12\big[ \sqrt{(1+\g_1)(1+\g_2)} + \sqrt{(1-\g_1)(1-\g_2)} \big].
\ee
If all eigenvalues of either \(\Gamma_1\) or \(\Gamma_2\) are \(\pm 1\), the corresponding density matrix is pure.
Using the fidelity between a pure state $|\psi\rag$ and a general state $\r$, $F(|\psi\rag,\r)=\sqrt{\lag\psi|\r|\psi\rag}$, as well as (\ref{trrG1rG2}), we obtain the fidelity
\be \label{Fpure}
F(\r_{\G_1},\r_{\G_2}) = \Big( \det\f{1+\G_1\G_2}{2} \Big)^{1/4}.
\ee

We next consider the more general case in which some, but not all, of the eigenvalues of one of the correlation matrices, $\Gamma_1 = R$ and $\Gamma_2 = S$, say $R$ without loss of generality, are equal to $\pm1$.
The correlation matrix $R$ has a total of $2\ell=2(x+y)$ eigenvalues, with $2x$ of them equal to $\pm1$ and the remaining $2y$ of them equal to $\pm r_j$, $j=1,2,\cdots,y$.
In the canonical basis of $R$, there is
\be
R = \bigg( \ba{cc} R_X & \\ & R_Y \ea \bigg), ~~
S = \bigg( \ba{cc} S_X & S_{XY} \\ S_{YX} & S_Y \ea \bigg),
\ee
with
\be
R_X = \bigoplus_{j=1}^x \bigg( \ba{cc} & \ii \\ -\ii & \ea \bigg), ~~
R_Y = \bigoplus_{j=1}^y \bigg( \ba{cc} & \ii r_j \\ -\ii r_j & \ea \bigg).
\ee
In fact, the following calculations in this section do not depend on the explicit forms of $R_X$ and $R_Y$, only the condition $\det R_X = 1$ is required.

The correlation matrix $R_X$ corresponds to a pure state in $\cH_X$ and we label it as $|1\rag$.
The density matrix corresponding to the correlation matrix $R$ is given by
\be
\r_R = ( |1\rag\lag1| ) \otimes \r_{R_Y},
\ee
from which we get
\be
\sqrt{\r_R} = ( |1\rag\lag1| ) \otimes \sqrt{\r_{R_Y}}.
\ee
We then obtain
\be
\sqrt{\r_R} \r_S \sqrt{\r_R} = ( |1\rag\lag1| ) \otimes ( \sqrt{\r_{R_Y}} \td \s \sqrt{\r_{R_Y}} ),
\ee
where the operator $\td \s$ acting on $\cH_Y$ is defined as
\be
\td \s = \tr_X \{ [ ( |1\rag\lag1| ) \otimes I_Y ] \r_S \},
\ee
with $I_Y$ denoting the identity operator on $\cH_Y$.
Both operators $( |1\rag\lag1| ) \otimes \f{I_Y}{2^y}$ and $\r_S$ are Gaussian density matrices, and from (\ref{rG1rG2}) we know that $\td \s$ is also a Gaussian density matrix.
We define the normalized density matrix
\be
\s = \f{\td\s}{\tr\td\s}.
\ee
Using (\ref{trrG1rG2}), we get the normalization factor
\be
\tr\td\s = \sqrt{\det\f{1+R_X S_X}{2}}.
\ee
From (\ref{G1timesG2}), we find that the correlation matrix corresponding to the density matrix $\s=\r_{\td S_Y}$
\be
\td S_Y = S_Y-S_{YX} R_X(1+S_X R_X)^{-1}S_{XY}.
\ee
Finally, we obtain the formula for the fidelity
\be \label{FrRrS}
F(\r_R,\r_S)= \Big(\det{\frac{1+R_X S_X}{2}}\Big)^{1/4} F(\r_{\td S_Y},\rho_{R_Y}),
\ee
which reduces the fidelity between two Gaussian states in a higher dimension to the fidelity between two other Gaussian states in a lower dimension.

\begin{figure}[tp]
  \centering
  \includegraphics[width=0.5\textwidth]{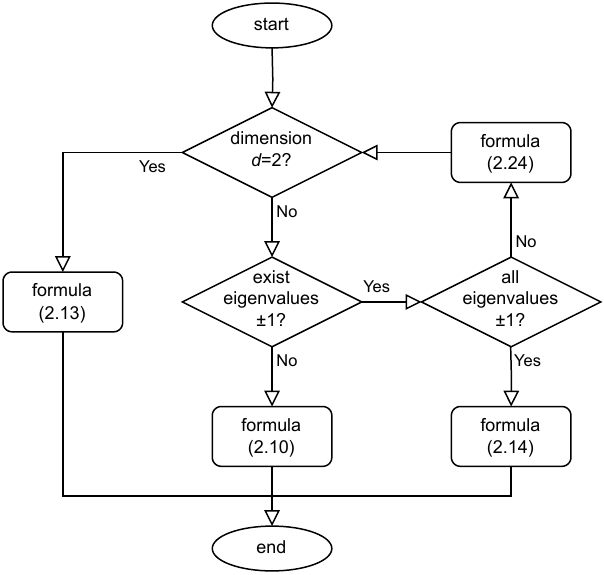}\\
  \caption{The algorithm for calculating the fidelity between two Gaussian states.}
  \label{FigureAlgorithm}
\end{figure}

Using the above formulas, we can calculate the fidelity between two Gaussian states efficiently as shown in the flowchart figure~\ref{FigureAlgorithm}.
If the Hilbert space dimension $d=2$, we simply use the formula (\ref{Fdeq2}).
If neither correlation matrix has eigenvalues $\pm 1$, we can use the formula (\ref{FrG1rG2II}), without concern for any divergence.
If one of the correlation matrices has all its eigenvalues equal to $\pm 1$, the corresponding state is pure and we calculate the fidelity using the formula (\ref{Fpure}).
When some of the eigenvalues of one correlation matrix are $\pm 1$, we use the formula (\ref{FrRrS}), which reduces the fidelity calculation in a higher-dimensional Hilbert space to that in a lower-dimension.
The algorithm can be applied iteratively and efficiently produces the numerical fidelity between two Gaussian states.

\section{Transverse field Ising chain}\label{sectionIsing}

In this section, we first briefly review the local conserved charges in the Ising chain. More details can be found in \cite{Grady:1982foy,Prosen:1998uvt}, especially in \cite{Fagotti:2013jzu}. Then we compute the average subsystem distance.

\subsection{Local conserved charges}\label{sectionLocal}

The Hamiltonian of transverse field Ising chain takes the form
\begin{equation}
    \label{hamiltonian}
    H =- \f12 \sum_{j=1}^L ( \sigma_j^x \sigma_{j+1}^x + h \sigma_j^z ),
\end{equation}
where we choose periodic boundary conditions (PBC) $\sigma_{L+1}^x=\sigma_1^x$.
After the Jordan-Wigner transformation and Bogoliubov transformation, the Hamiltonian can be diagonalized as \cite{Lieb:1961fr,Katsura:1962hqz,Pfeuty:1970ayt}
\begin{equation}
H = \frac{1+P}{2} H_\NS + \frac{1-P}{2} H_\R, ~~
H_\NS = \sum_{k \in \NS} \ce_k \Big(c_k^\dagger c_k-\frac{1}{2}\Big), ~~
H_\R = \sum_{k \in \R}  \ce_k \Big(c_k^\dagger c_k-\frac{1}{2}\Big),
\end{equation}
where $c_k$ and $c^\dag_k$ are modes of complex fermions. The energy of the elementary excitation is
\be
\ve_k = \sqrt{h^2-2h\cos{\f{2\pi k}{L} +1 }}.
\ee
In the Neveu-Schwarz (NS) sector, the momentum $k$ takes values in the set
\be
k \in  \Big\{ \f12, \f32, \cdots, L-\f12 \Big\},
\ee
and in the Ramond (R) sector, the momentum $k$ takes values in the set
\be
k \in  \{ 0,1, \cdots, L-1 \}.
\ee
The states with an even number of elementary excitations in NS sector and those with odd number of elementary excitations in R sector belong to the spectrum of the model with PBC.
The total momentum is a conserved quantity,
\be
K = {\rm mod} \Big( \sum_{k} k c^\dag_k c_k, L \Big),
\ee
whose eigenvalues lie in the set $K \in \{0,1,\cdots,L-1\}$.
The parity operator $P$ is also conserved,
\begin{equation}
P = \prod_{j=1}^L \sigma_j^z = \exp{(\pm \pi \ii \sum_{k} c_k^\dagger c_k)},
\end{equation}
whose eigenvalues are $\pm1$.

In the following sections, we calculate the average subsystem Bures distance between neighboring eigenstates in the transverse field Ising chain.
We consider the whole spectrum, the sector with fixed parity $P$, the sector with fixed momentum $K$, as well as the sector with both fixed $P$ and $K$.
In any case, there are issues of energy degeneracy which we will lift using the local conserved charges \cite{Fagotti:2013jzu}
\be \label{conservedcharge}
Q_n^{+} = \sum_{k} \cos\f{n k \pi}{L} \varepsilon_k \Big( c_k^\dagger c_k - \f12 \Big), ~~
Q_n^{-} = \sum_{k} \sin\f{(n+1) k \pi}{L} \Big( c_k^\dagger c_k - \f12 \Big), ~~
n=0,1,\cdots,\Big\lfloor\f{L}{2}\Big\rfloor.
\ee
For $n=0$, $Q_0^{+}$ equals either $H_{\text{NS}}$ or $H_{\text{R}}$, and the sum over momenta $k$ in equation (\ref{conservedcharge}) depends on the symmetry sector (NS or R) under consideration. The local terms in $Q_n^\pm$ only involve $n+2$ neighboring sites of the periodic Ising chain. We relabel the charges as
\be \label{Qm}
Q_m = \lt\{
\ba{ll}
Q_{m/2}^+     & m \in {\rm~even} \\
Q_{(m-1)/2}^- & m \in {\rm~odd}
\ea
\rt.\!\!\!.
\ee
where $m=0,1,\cdots,L-1$.
The states in the spectrum are primarily sorted by their energy $Q_0$.
When there is degeneracy for $Q_0$, we use $Q_1$ to sort degenerate states.
When there is degeneracy for $(Q_0,Q_1)$, we use $Q_2$ to sort degenerate states.
This hierarchical sorting process continues iteratively until either (i) all degeneracies are resolved, or (ii) the maximum sorting level $m=L-1$ is reached.

\begin{figure}[tp]
  \centering
  \includegraphics[width=0.5\textwidth]{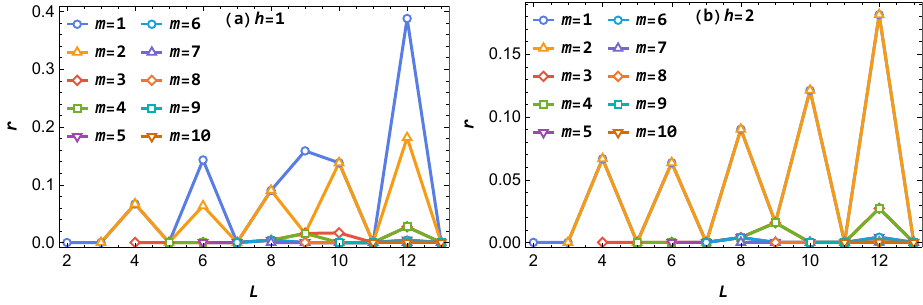}
  \caption{The ratio of degenerate pairs $r$ (\ref{defr}) in the whole spectrum after sorted using the local conserved charges $(Q_0,Q_1,\cdots,Q_m)$.}
  \label{FigureDegeneracy}
\end{figure}

The complete spectrum consists of $2^L$ eigenstates, giving $2^L-1$ adjacent state pairs.
For states sorted by the charge operators $(Q_0,Q_1,\cdots,Q_m)$, the degeneracy ratio $r$ is defined as
\be \label{defr}
r = \f{{\rm number~of~degenerate~pairs}}{2^L-1}.
\ee
Figure~\ref{FigureDegeneracy} displays the dependence of $r$ on the parameter $h$, $m$ and $L$.
We see that for fixed $h$ and $L$, the ratio of degenerate pairs $r$ goes to zero quickly with the increase of $m$.
Actually, the number of degeneracy can be always vanishing for some $m\leq L-1$. 
Notably, when $L$ is prime, sorting ($m=1$) suffices to eliminate all degeneracies, whereas for $L$ being multiples of 4, sorting ($m=L-1$) is required.

The same procedure can be applied to sectors with fixed $P$, $K$, or their combination $(P,K)$.
Compared to the full system, these subsystems show both a lower degeneracy ratio $r$ and require smaller values of $m$ to completely remove all degeneracies.
We do not show details here. The lift of all the degeneracies guarantees the uniqueness of the results in the following section.

\subsection{Average subsystem distance}\label{sectionDistance}

Having established the Bures distance calculation algorithm and resolved the eigenstate degeneracy issue, we now compute the average subsystem distance for both the full spectrum and specific symmetry sectors in the transverse field Ising chain.

\begin{figure}[t]
  \centering
  \includegraphics[width=0.235\textwidth]{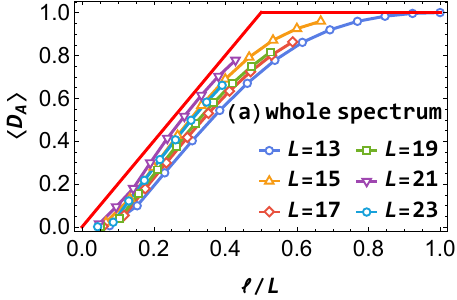} ~
  \includegraphics[width=0.235\textwidth]{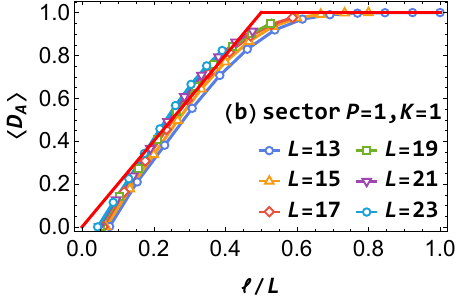} ~
  \includegraphics[width=0.235\textwidth]{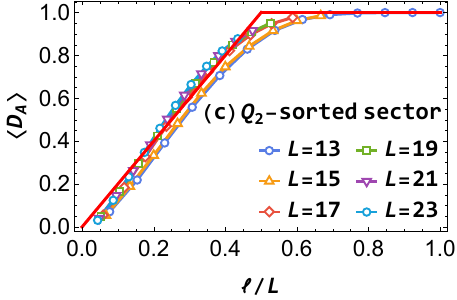} ~
  \includegraphics[width=0.235\textwidth]{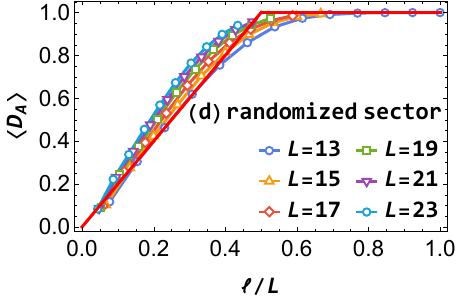} \\
  \includegraphics[width=0.235\textwidth]{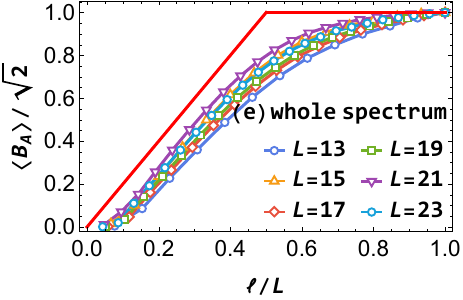} ~
  \includegraphics[width=0.235\textwidth]{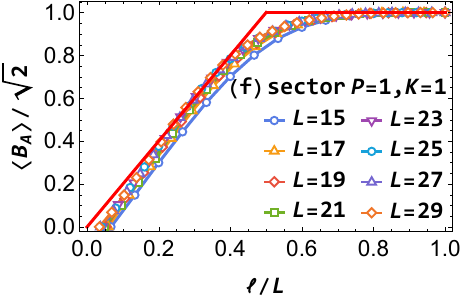} ~
  \includegraphics[width=0.235\textwidth]{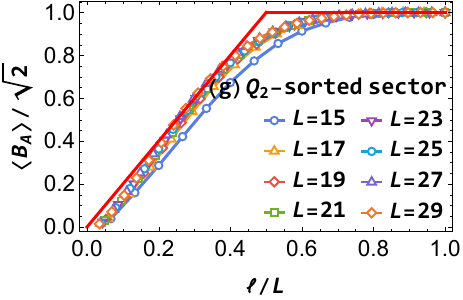} ~
  \includegraphics[width=0.235\textwidth]{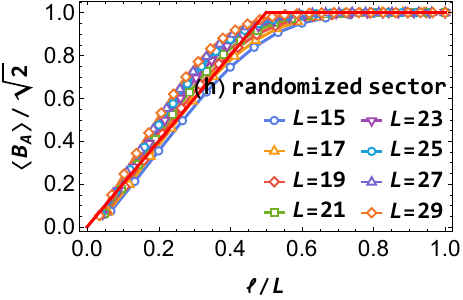}
  \caption{The average subsystem trace distance (top panels) and Bures distance (bottom panels) in the transverse field Ising chain. The 1st column is for the whole spectrum, the 2nd column is for the sector with fixed parity $P=1$ and momentum $K=1$, the 3rd colum is for the sector with fixed parity $P=1$ and momentum $K=1$ which is sorted firstly by $Q_2$ then by $Q_0,Q_1,Q_3,\cdots$, and the 4th column is for the sector with fixed parity $P=1$ and momentum $K=1$ in which the states are randomly permuted.
  The solid red lines are the function (\ref{universalfx}).
  In all the panels, we have set the transverse field $h=1$.}
  \label{FigureAD}
\end{figure}

\begin{figure}[t]
  \centering
  \includegraphics[width=0.99\textwidth]{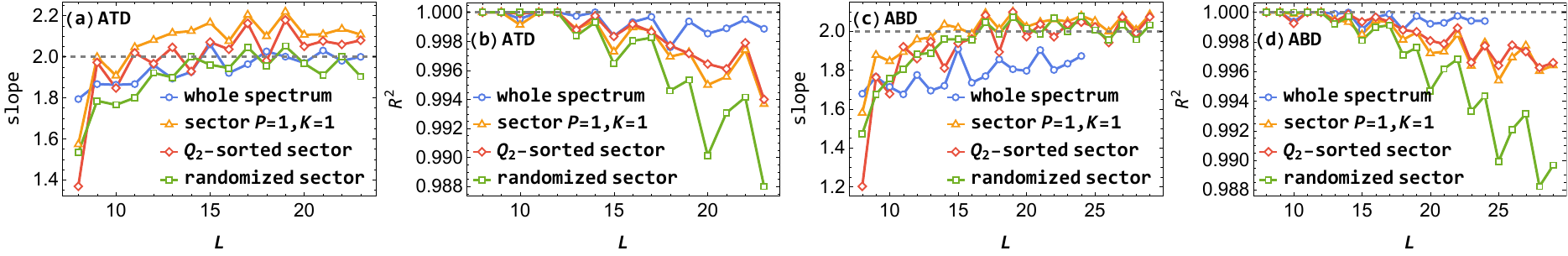}
  \caption{The fitted linear slopes for the average subsystem trace $\langle D_A \rangle$ (panel (a)) and the scaled Bures distances $\langle B_A \rangle/\sqrt{2}$ (panel (c)) are shown in figure~\ref{FigureAD} for $\lceil 0.2 L \rceil \leq \ell \leq \lfloor 0.4 L \rfloor$. The corresponding $R^2$ values, which quantify the accuracy of the linear fits, are shown in panels (b) and (d). As the total system size $L$ increases, all slopes approach approximately 2. Although $R^2$ appears to decrease with increasing $L$, its values remain very close to 1, indicating good fits.}
  \label{FigureSlope}
\end{figure}

We first consider the average subsystem trace distance and Bures distance in the whole spectrum.
The results are shown in the first column of figure~\ref{FigureAD}.
For subsystem sizes satisfying $\ell/L < 1/2$, the average distance exhibits an approximately linear dependence: $a \f{\ell}{L}+b$.
Within the range $ \lceil 0.2 L \rceil \leq \ell \leq \lfloor 0.4 L \rfloor$, we perform linear regression to extract the slope $a$, with the fitting results of the slope presented in figure~\ref{FigureSlope}.
Next, we examine the average subsystem distance in the sector with fixed paraity $P$ and momentum $K$, as shown in the second column of figure~\ref{FigureAD}. The fitted slope $a$ is also shown in figure~\ref{FigureSlope}.
These results are consistent with the conjectured universal slope of $2$ reported in \cite{Khasseh:2023kxw} and suggest a possible enhanced universal scaling behavior: both the average trace distance $\langle D_A \rangle$ and the scaled Bures distance $\langle B_A \rangle/\sqrt{2}$ follow a piecewise linear function
\be \label{universalfx}
f(x) = \lt\{
\ba{ll}
2x & 0 \leq x<\f12 \\
1  & \f12<x \leq 1
\ea
\rt.\!\!\!,
\ee
with possible deviation around $x=\f12$. This possible universal scaling is highlighted by the solid red lines in figure~\ref{FigureAD}, as well as in subsequent figure~\ref{FigureATDABTrandom} and figure~\ref{FigureXXZ} in the appendix.

\begin{figure}[t]
  \centering
  \includegraphics[width=0.235\textwidth]{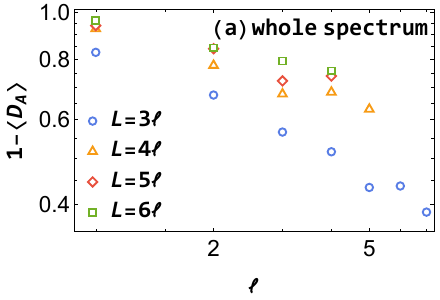} ~
  \includegraphics[width=0.235\textwidth]{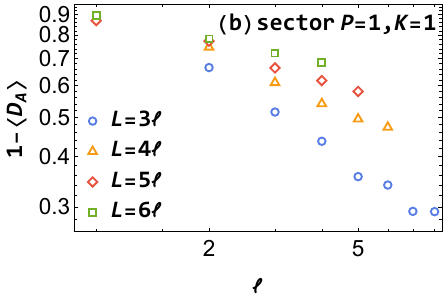} ~
  \includegraphics[width=0.235\textwidth]{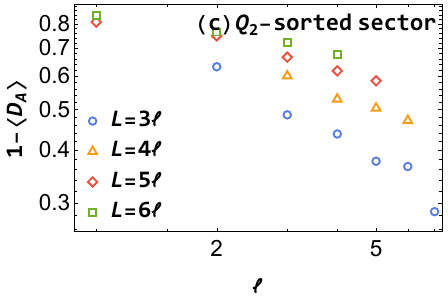} ~
  \includegraphics[width=0.235\textwidth]{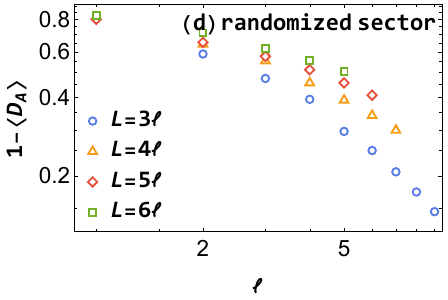} \\
  \includegraphics[width=0.235\textwidth]{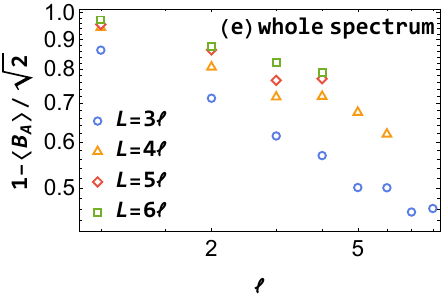} ~
  \includegraphics[width=0.235\textwidth]{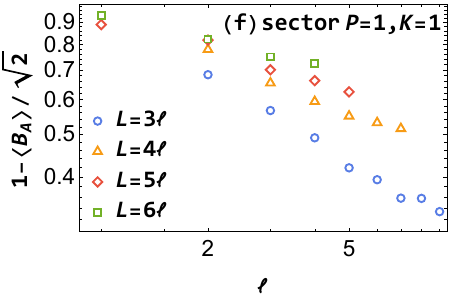} ~
  \includegraphics[width=0.235\textwidth]{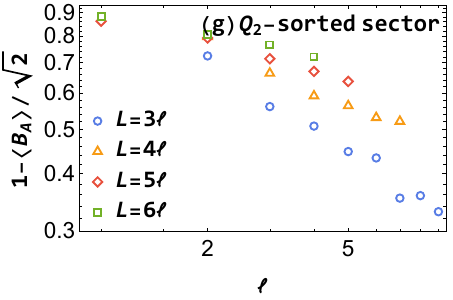} ~
  \includegraphics[width=0.235\textwidth]{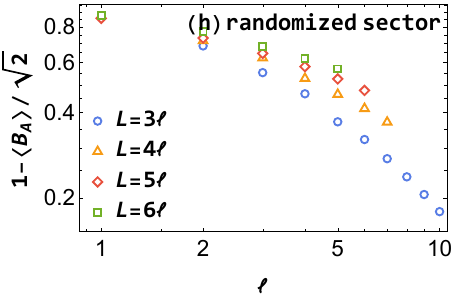}
  \caption{The average subsystem trace distance (top panels) and Bures distance (bottom panels) in a transverse field Ising chain. The spectrum sectors are identical to those in figure~\ref{FigureAD}, and we have also set $h=1$.}
  \label{FigureAD4x}
\end{figure}

To test the applicability of the conjectured function (\ref{universalfx}), we take one of the other charges, say $Q_2$, as the Hamiltonian. Although the simultaneous eigenstates of all charges remain unchanged, their ordering should be reassigned. We first sort the states by the Hamiltonian $Q_2$, then by $Q_0$ for degeneracies, then by $Q_1$ for any remaining degeneracies, then by $Q_3$, and so on. The resulting average subsystem trace and Bures distances are shown in the third column of figure~\ref{FigureAD}.

The similarity of the second and third columns in figure~\ref{FigureAD} suggests that the ordering of the eigenstates may be irrelevant. To test this, we randomly permute the states and display the resulting average subsystem trace and Bures distances in the fourth column of figure~\ref{FigureAD}; the corresponding fitted slopes and $R^2$ of the fits are shown in figure~\ref{FigureSlope}. Although the fitted slopes are approximately 2 and the $R^2$ values are very close to 1, the average subsystem trace and Bures distances show slight deviations from the universal function in Eq.~\eqref{universalfx}, particularly for the randomly permuted sector. As shown in the fourth column of figure~\ref{FigureAD4x}, the distances for a fixed \(x = \frac{\ell}{L} < \frac{1}{2}\) appear to approach \(\langle D_A \rangle \to 1\) and \(\langle B_A \rangle \to \sqrt{2}\) in the scaling limit.
For this case we would have both the average trace distance $\langle D_A \rangle$ and the scaled Bures distance $\langle B_A \rangle/\sqrt{2}$ approach
\be \label{universalgx}
g(x) = \lt\{
\ba{ll}
0 & x \to 0 \\
1 & 0<x<1
\ea
\rt.\!\!\!.
\ee
It is less clear for other panels of figure~\ref{FigureAD4x}. Whether the results in the first, second, and third columns of figure~\ref{FigureAD} approach the function (\ref{universalfx}) or (\ref{universalgx}) as the system size increases remains an open question.

\section{Dirac SYK$_2$ model} \label{sectionSYK2}

The SYK models \cite{Sachdev:1992fk,Kitaev:2015qjp,Sachdev:2015efa,Polchinski:2016xgd,Maldacena:2016hyu} represent a family of solvable quantum many-body models of randomly interacting Majorana fermions. In its \( q \)-interacting form with \( q>2 \), the SYK models exhibit maximal chaos and serve as holographic duals to quantum gravity systems that contain black holes. In this section, we consider the average subsystem distance in the Dirac fermion formulation of the SYK model with $q=2$, called the Dirac SYK$_2$ model.

The Dirac SYK$_2$ model is defined by the quadratic Hamiltonian
\be \label{HDSY2}
H = \sum_{j_1,j_2=1}^L A_{j_1j_2} a^\dag_{j_1} a_{j_2},
\ee
where \( A \) is an \( L \times L \) Hermitian matrix. Its matrix elements are independently drawn from Gaussian distributions: the diagonal elements \(A_{jj}\) have mean zero and variance \(2/L\), while the real and imaginary parts of the off-diagonal elements \(A_{j_1 j_2}\) with \(j_1 < j_2\) each have mean zero and variance \(1/L\).
The Hamiltonian (\ref{HDSY2}) resembles that of the Anderson model, but differs fundamentally in its structure. The Anderson model describes a lattice with local hopping, whereas the Dirac SYK$_2$ model is a zero-dimensional, all-to-all random hopping model with no spatial geometry.
The Hamiltonian conserves the total particle number
\be
N = \sum_{j=1}^L a^\dag_j a_j.
\ee

The Dirac SYK$_2$ model features a quantum-chaotic quadratic Hamiltonian, meaning its single-particle eigenstates exhibit quantum chaos; that is, its single-particle spectrum follows that of random matrix theory. Its entanglement entropy has been studied extensively \cite{Liu:2017kfa,Zhang:2020kia,Lydzba:2020qfx,Lydzba:2021hml}. In the thermodynamic limit, the average entanglement entropy for such random quadratic Hamiltonians equals that of pure random fermionic Gaussian states \cite{Bianchi:2021lnp}. This result is close to but differs slightly from the average entanglement entropy in integrable models \cite{Vidmar:2017uux,Vidmar:2018rqk,Hackl:2018tyl,Jafarizadeh:2019xxc,LeBlond:2019eoe}.

\begin{figure}[t]
  \centering
  \includegraphics[width=0.235\textwidth]{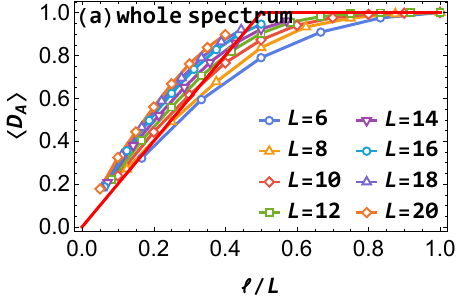} ~
  \includegraphics[width=0.235\textwidth]{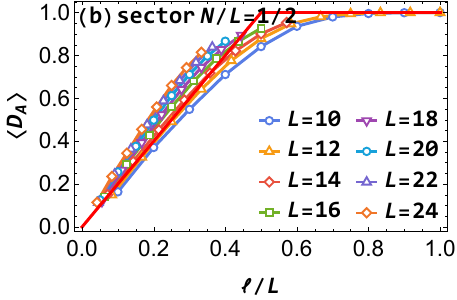} ~
  \includegraphics[width=0.235\textwidth]{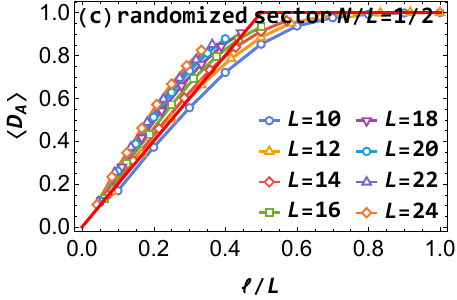} \\
  \includegraphics[width=0.235\textwidth]{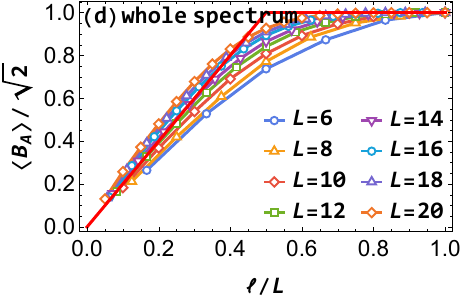} ~
  \includegraphics[width=0.235\textwidth]{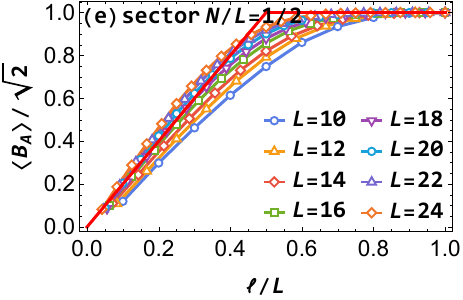} ~
  \includegraphics[width=0.235\textwidth]{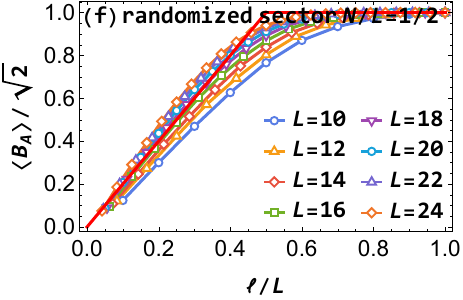}
  \caption{Average subsystem distance in the Dirac SYK\(_2\) model. The first column shows results for neighboring eigenstates from the full spectrum. The second and third columns show results for the symmetry-resolved sector with fixed filling fraction \(N/L = 1/2\); the second column corresponds to neighboring eigenstates, while the third column corresponds to random pairs of eigenstates. To reduce statistical errors, averages are taken over both the relevant pairs of eigenstates and 32 independent realizations of the random Hamiltonian. The red lines in all panels represent the function given in equation~\eqref{universalfx}.}
  \label{FigureSYK2ATDABD}
\end{figure}

\begin{figure}[t]
  \centering
  \includegraphics[width=0.235\textwidth]{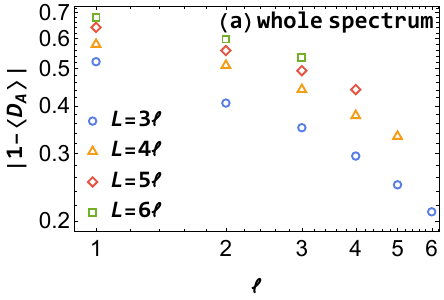} ~
  \includegraphics[width=0.235\textwidth]{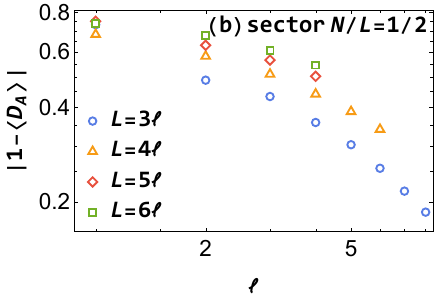} ~
  \includegraphics[width=0.235\textwidth]{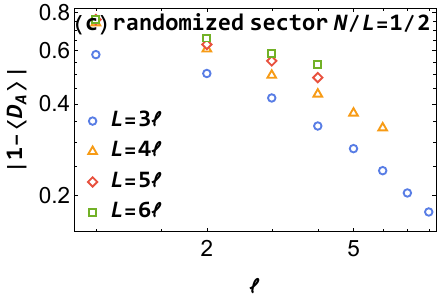} \\
  \includegraphics[width=0.235\textwidth]{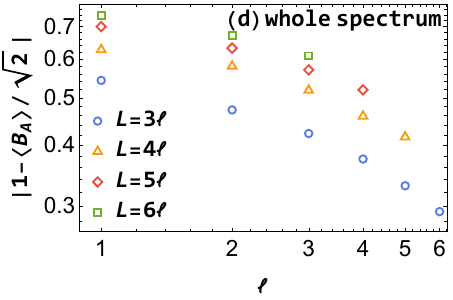} ~
  \includegraphics[width=0.235\textwidth]{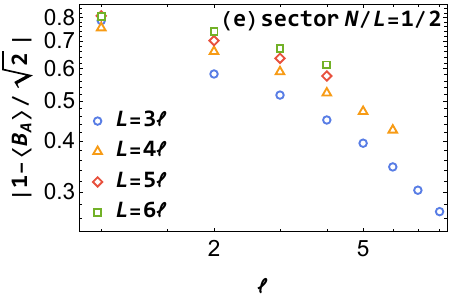} ~
  \includegraphics[width=0.235\textwidth]{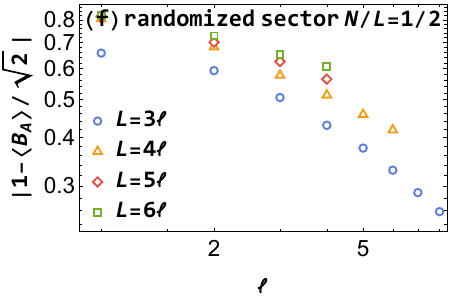}
  \caption{Average subsystem distance in the Dirac SYK$_2$ model. Each panel here corresponds to its counterpart in figure~\ref{FigureSYK2ATDABD}.}
  \label{FigureSYK2ATDABDx}
\end{figure}

We present the average subsystem distance for the Dirac SYK$_2$ model in figures~\ref{FigureSYK2ATDABD} and \ref{FigureSYK2ATDABDx}. The results reveal several notable features:
\begin{itemize}
    \item In panels (a) of both figures, the average trace distance \(\langle D_A \rangle\) appears to approach 1 for a fixed subsystem fraction \(x=\ell/L \in (0, 1/2)\) in the scaling limit. Panels (d) similarly show the average Bures distance \(\langle B_A \rangle\) tending asymptotically toward \(\sqrt{2}\). This is consistent with the inequalities (\ref{bounds}), since the convergence \(\langle D_A \rangle \to 1\) necessarily implies \(\langle B_A \rangle \to \sqrt{2}\). We have both $\langle D_A \rangle$ and $\langle B_A \rangle/\sqrt{2}$ approach (\ref{universalgx}) in the scaling limit.
    \item The results in the second and third columns are nearly identical, indicating that eigenstate ordering is not a significant factor here.
    \item The results for the sector with fixed filling fraction \(\f{N}{L} = \f12\) are similar to those for the full spectrum. In panel (b), it is evident that \(\lag D_A \rag \to 1\) in the scaling limit. The corresponding behavior of \(\lag B_A \rag\) in panel (e) is less distinct; however, the convergence \(\langle D_A \rangle \to 1\) necessarily implies \(\langle B_A \rangle \to \sqrt{2}\).
\end{itemize}

\section{Random pure Gaussian states} \label{sectionRandom}

The Dirac SYK$_2$ model is a quantum chaotic quadratic model, and the average entanglement entropy in quantum chaotic quadratic models has been shown to be the same as that of random pure Gaussian states \cite{Bianchi:2021lnp}. To verify the average subsystem distances in the Dirac SYK$_2$ model from the previous section, we calculate the average subsystem trace and Bures distances between random pure Gaussian states in this section.

To define a Gaussian state $\rho_\Gamma$ of $L$ qubits, we need a $2L \times 2L$ purely imaginary correlation matrix $\Gamma$. For pure Gaussian states, $\det \Gamma = 1$.
A random pure Gaussian state $\rho_\Gamma$ (size $2^L \times 2^L$) can be obtained through the following steps:
\begin{itemize}
  \item Generate a random real orthogonal matrix $U$ (size $2L \times 2L$). This matrix is drawn from the circular real ensemble based on the Haar measure.
  \item Calculate the $2L \times 2L$ correlation matrix using
   \be
   \Gamma = U \Big( \bigoplus_{j=1}^L \sigma^y \Big) U^T, ~~
   \sigma^y = \bigg( \ba{cc} 0 & -\ii \\ \ii & 0 \ea \bigg).
   \ee
\end{itemize}

\begin{figure}[t]
  \centering
  \includegraphics[width=0.235\textwidth]{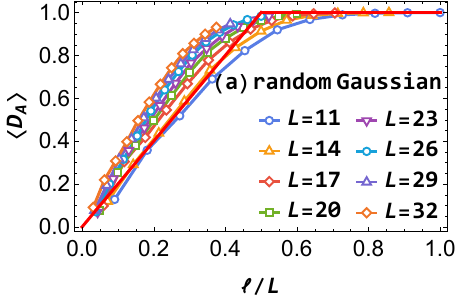} ~
  \includegraphics[width=0.235\textwidth]{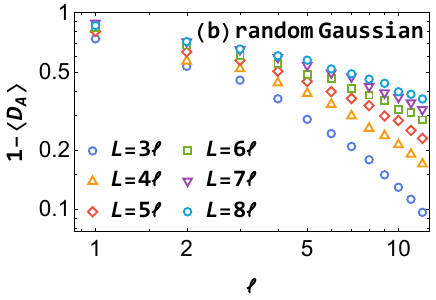} ~
  \includegraphics[width=0.235\textwidth]{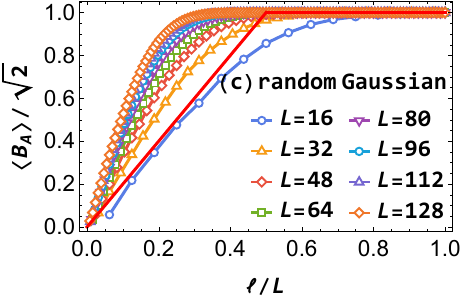} ~
  \includegraphics[width=0.235\textwidth]{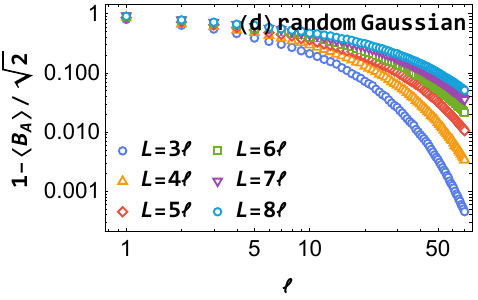}
  \caption{The average subsystem trace distance (1st and 2nd panels) and Bures distance (3rd and 4th panels) between random Gaussian states. For each value of the system size $L$, we generate 32 random pure Gaussian states, and the average is taken over all 496 possible pairs. The solid red lines are the function (\ref{universalfx}).}
  \label{FigureATDABTrandom}
\end{figure}

We show the average subsystem trace distance and Bures distance between random Gaussian states in figure~\ref{FigureATDABTrandom}.
In the first and third panels, no linear growth is observed for large system sizes.
In the second and fourth panels, we see that for a fixed ratio $x=\f{\ell}{L}\in(0,\f12)$, both the average subsystem trace distance and the Bures distance approach their respective maximal values as the system size increases, tending toward the universal function (\ref{universalgx}) in the scaling limit.

\begin{figure}[t]
  \centering
  \includegraphics[width=0.5\textwidth]{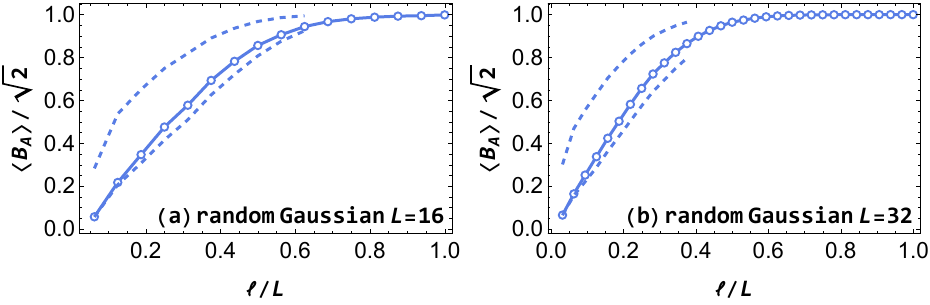}
  \caption{Examples of the average subsystem Bures distance between random Gaussian states (joined symbols) and the upper and lower bounds derived from the average subsystem trace distance (dashed lines).}
  \label{FigureATBDrandom}
\end{figure}

For random Gaussian states, figure~\ref{FigureATBDrandom} displays examples of the average subsystem Bures distance alongside its upper and lower bounds (\ref{bounds}) from the average subsystem trace distance. The Bures distance is close to its lower bound, yet neither bound is saturated.

In summary, the average subsystem distances panels (a) and (d) of figure~\ref{FigureSYK2ATDABD} for the full spectrum of the Dirac SYK$_2$ model matches the result for random pure Gaussian states shown in figure~\ref{FigureATBDrandom}. In contrast, the average subsystem distance in the symmetry-resolved sector \(N=1\) is markedly different. The results for the average subsystem distance presented here are consistent with those for the average entanglement entropy in \cite{Lydzba:2020qfx,Lydzba:2021hml,Bianchi:2021lnp,Bianchi:2021aui}. This is expected, as the eigenstates of quantum-chaotic quadratic Hamiltonians are just random pure Gaussian states, in contrast to those of integrable Hamiltonians.

\section{Conclusion and discussions}\label{sectionDiscussion}

We developed an efficient algorithm for computing the Bures distance between fermionic Gaussian states, which allows evaluation of the average subsystem Bures distance at larger system sizes. The method was applied to eigenstates of the spin-1/2 transverse-field Ising chain, where degeneracies are resolved by constructing simultaneous eigenstates of all local conserved charges and ordering states accordingly, and to the Dirac SYK$_2$ model, as well as to random pure fermionic Gaussian states.
The average subsystem distances for the whole spectrum in the transverse-field Ising model agree with the conjectured linear growth of the average subsystem distance with subsystem size in quadratic integrable systems. In contrast, for the Dirac SYK$_2$ model and random pure Gaussian states, the average subsystem distance does not exhibit such linear growth.

\begin{figure}[t]
  \centering
  \includegraphics[width=0.235\textwidth]{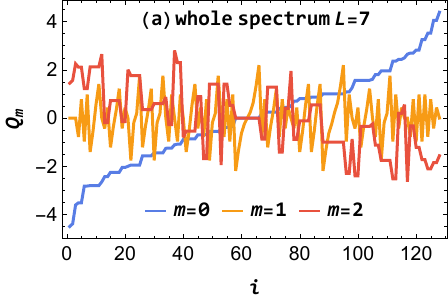} ~
  \includegraphics[width=0.235\textwidth]{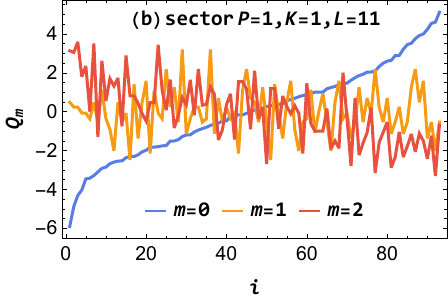} ~
  \includegraphics[width=0.235\textwidth]{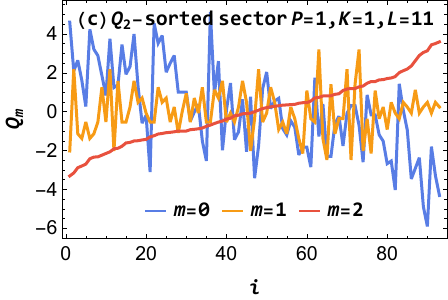} ~
  \includegraphics[width=0.235\textwidth]{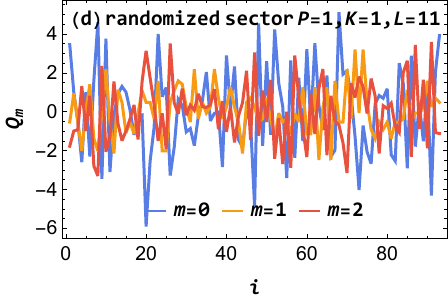}
  \caption{Charges $Q_m$ ($m=0,1,2$) shown for: the whole spectrum (1st panel); the sector with fixed $P=1$ and $K=1$ (2nd panel); the $Q_2$-sorted sector ($P=1$, $K=1$) (3rd panel); and the randomly permuted sector ($P=1$, $K=1$) (4th panel) in the transverse field Ising chain. Horizontal coordinate $i$ denotes eigenstate index. All panels use $h=1$.}
  \label{FigureCharges}
\end{figure}

In chaotic systems, it has been conjectured \cite{Khasseh:2023kxw} that the average subsystem trace distance vanishes in the scaling limit $L\to+\inf$ with fixed ratio $x=\f{\ell}{L}<\f12$, the same as the average subsystem trace distance between random states \cite{Kudler-Flam:2021rpr,Kudler-Flam:2021alo,deMiranda:2022rze}.
This aligns with ETH, which posits that the expectation values of local operators vary smoothly with energy and, in turn, guarantees that adjacent energy eigenstates have nearly identical reduced density matrices; hence both the pairwise and the average subsystem trace distances are small.
Integrable systems, by contrast, possess an extensive set of local conserved charges. The notion of generalized eigenstate thermalization \cite{Vidmar:2016laa,Cassidy:2011smq}, corroborated by recent experiments \cite{Wang:2022ufp}, accounts for this additional structure. Representative conserved charges across the full spectrum and within a sector with fixed momentum $P$ and parity $K$ are displayed in figure~\ref{FigureCharges}.
Because eigenstates adjacent in energy can differ appreciably in the other local charges, their RDMs, determined by the full set of conserved quantities, are no longer guaranteed to be close. Consequently, the average subsystem distance does not tend to zero in the same scaling limit.

\begin{figure}[t]
  \centering
  \includegraphics[width=0.235\textwidth]{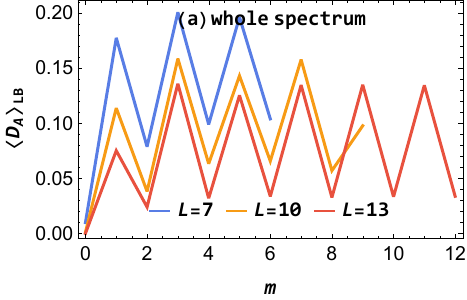} ~
  \includegraphics[width=0.235\textwidth]{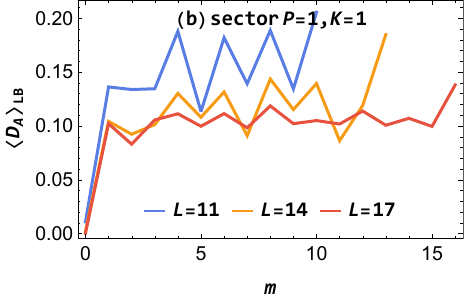} ~
  \includegraphics[width=0.235\textwidth]{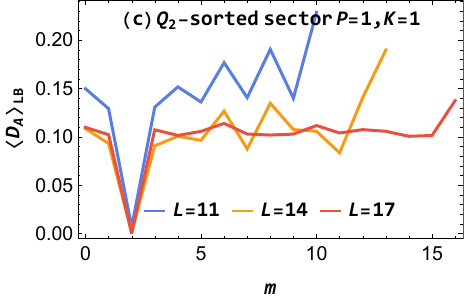} ~
  \includegraphics[width=0.235\textwidth]{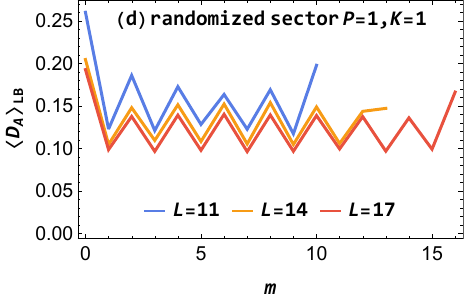}
  \caption{Lower bound of average subsystem trace distance from local conserved charge \(Q_m\) expectation values in the transverse field Ising chain (\(h=1\)): (a) the whole spectrum, (b) sector with fixed \(P=1,K=1\), (c) \(Q_2\)-sorted sector with \(P=1,K=1\), and (d) randomly permuted sector with \(P=1,K=1\).}
  \label{FigureChargesx}
\end{figure}

The trace distance \(D(\r,\s)\) between two states \(\r\) and \(\s\) is bounded by the difference in expectation values
\be
| \lag Q \rag_\r - \lag Q \rag_\s | \leq 2 s_Q D(\r,\s),
\ee
where \(s_Q\) is the largest singular value of operator \(Q\). The conserved charges \(Q_m\) are sums of order \(L\) terms, each being the product of Pauli matrices for \(\frac{m}{2}+2\) neighboring sites, leading to \(s_{Q_m} \sim O(L)\). For a subsystem \(A\) of size \(\ell \gtrsim \frac{m}{2}+2\), the average subsystem trace distance has a lower bound
\be
\lag D_A \rag \gtrsim \lag D_A \rag_{\rm{LB}} \equiv \frac{1}{(d-1)L} \sum_{i=1}^{d-1} | \lag Q_m \rag_i - \lag Q_m \rag_{i+1} |.
\ee
In cases shown in figure~\ref{FigureCharges}, the lower bound of the average subsystem trace distance is displayed in figure~\ref{FigureChargesx}. As the trace distance increases monotonically with subsystem size, the lower bound tightens for smaller \(m\), supporting the previous qualitative argument.

Although the average subsystem distance in chaotic and integrable systems is obviously distinct, for $x<1/2$ the former vanishes while the latter is finite. While the reason for the difference is clear, the precise behavior of the average subsystem distance in integrable systems remains unknown, possibly corresponding to the function (\ref{universalfx}) or (\ref{universalgx}). This remains to be further investigated.

To further test the universality of average subsystem distance in integrable systems, we compute the average subsystem trace and Bures distances in spin-1/2 XXZ chain in appendix~\ref{appXXZ}. While the average subsystem trace distance has been calculated for a sector with fixed momentum in \cite{Khasseh:2023kxw}, the limited system sizes precluded an accurate slope extraction and the ordering of states remained uncertain. In appendix~\ref{appXXZ}, we instead fix both momentum and magnetization, allowing us to reach larger system sizes and to evaluate the average subsystem distance unambiguously by computing the distance between every pair of states in the sector. The current results favor a slope smaller than 2; however, the precise value remains to be determined. In the XXZ chain and more general integrable systems, the eigenstates are generally non-Gaussian, and efficient algorithms are needed to calculate the average subsystem distance.

If the universal slope of 2 exists for the Ising chain and even more general quadratic integrable systems, it raises the question of how to interpret it. We briefly discuss this open question in appendix~\ref{appSlo}. One possible interpretation involves the average difference in mode numbers between neighboring states within the spectrum. However, our analysis demonstrates that this particular explanation is inaccurate.

\section*{Acknowledgements}

We thank Markus Heyl and Reyhaneh Khasseh for helpful comments and discussions.
MAR is grateful to ICTP for hosting their visit in 2025, where a portion of this work was completed.
ZG and JZ acknowledge support from the National Natural Science Foundation of China (NSFC) grant number 12205217 and Tianjin University Self-Innovation Fund Extreme Basic Research Project grant number 2025XJ21-0007.
MAR thanks CNPq and FAPERJ (grant number E-26/210.062/2023) for partial support.
Numerical calculations for this study were performed at high performance cluster at Center for Joint Quantum Studies (HPC-CJQS) of Tianjin University.

\appendix

\section{Sector with fixed particle number in Dirac SYK$_2$ model} \label{appSYK2}

\begin{figure}[t]
  \centering
  \includegraphics[width=0.235\textwidth]{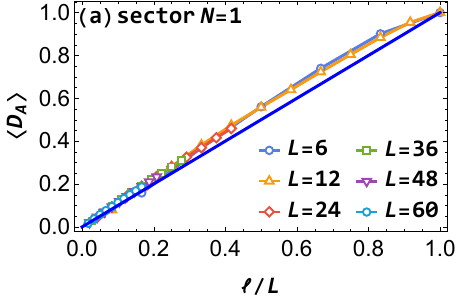} ~
  \includegraphics[width=0.235\textwidth]{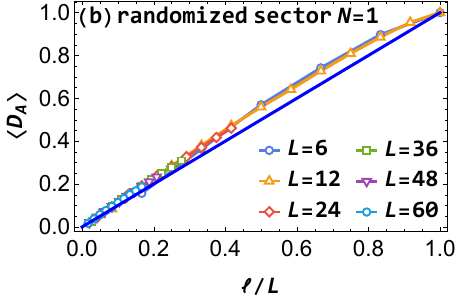} ~
  \includegraphics[width=0.235\textwidth]{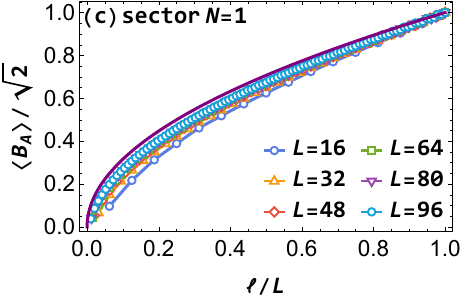} ~
  \includegraphics[width=0.235\textwidth]{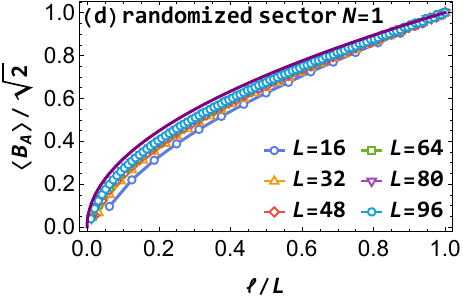} \\
  \includegraphics[width=0.235\textwidth]{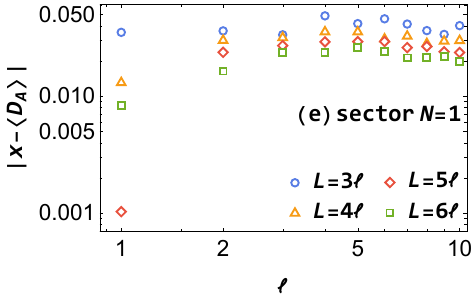} ~
  \includegraphics[width=0.235\textwidth]{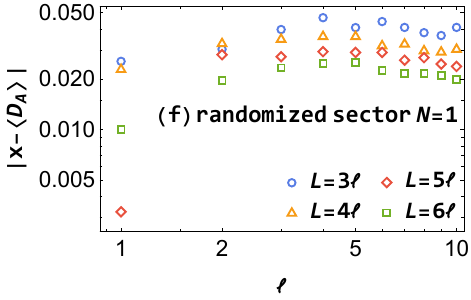} ~
  \includegraphics[width=0.235\textwidth]{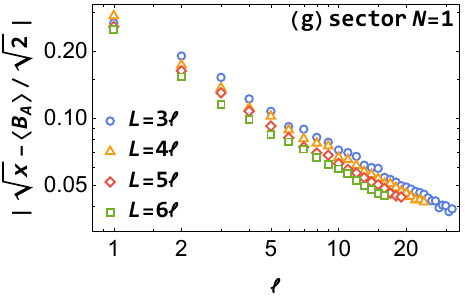} ~
  \includegraphics[width=0.235\textwidth]{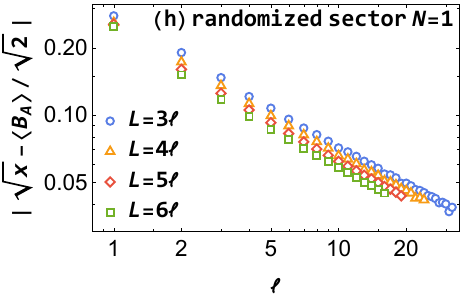}
  \caption{Average subsystem distance in the symmetry-resolved sector with particle number \(N=1\) of the Dirac SYK\(_2\) model. The first and third columns correspond to neighboring eigenstates, while the second and fourth columns correspond to all pairs of eigenstates. Averages are performed over both the relevant pairs of eigenstates and 32 independent realizations of the random Hamiltonian. The blue solid lines in panels (a) and (b) represent \(x = \ell/L\), and the purple lines in panels (c) and (d) represent the function \(\sqrt{x}\).}
  \label{FigureSYK2ATDABDapp}
\end{figure}

We calculate the average subsystem distance in the sector with particle number \(N=1\) of the Dirac SYK\(_2\) model and present the results in figure~\ref{FigureSYK2ATDABDapp}. In the scaling limit, the average subsystem trace distance \(\lag D_A \rag\) as a function of the ratio \(x = \f{\ell}{L}\) is slightly larger than \(x\), while the average subsystem Bures distance satisfies \(\lag B_A \rag \to \sqrt{2x}\). This behavior is also consistent with the inequalities (\ref{bounds}). The average subsystem distance in the sector with fixed particle number \(N=1\) differs markedly from those in the full spectrum and in the sector with fixed filling fraction \(\f{N}{L} = \f12\) discussed in section~\ref{sectionSYK2}.

We observe that the maximum accessible system size for the SYK$_2$ model is constrained predominantly by the requirement to average over numerous disorder realizations to yield smooth statistical results, rather than by the intrinsic scaling bottleneck of the Bures distance algorithm itself. For the $N=1$ sector of a fixed Hamiltonian, the algorithm maintains excellent performance for system sizes up to $L\sim100$.

\section{XXZ chain}\label{appXXZ}

\begin{figure}[t]
  \centering
  \includegraphics[width=0.235\textwidth]{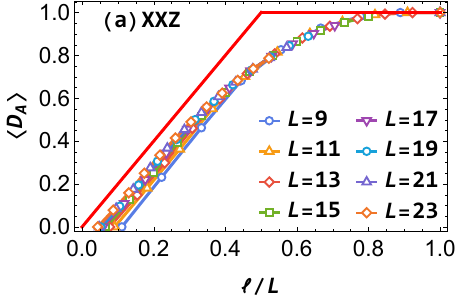} ~
  \includegraphics[width=0.235\textwidth]{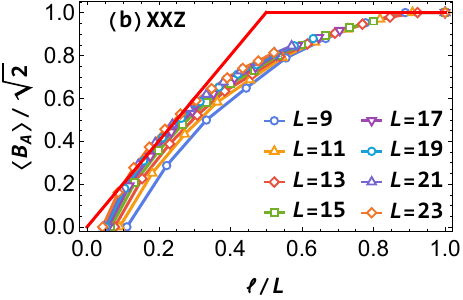} ~
  \includegraphics[width=0.235\textwidth]{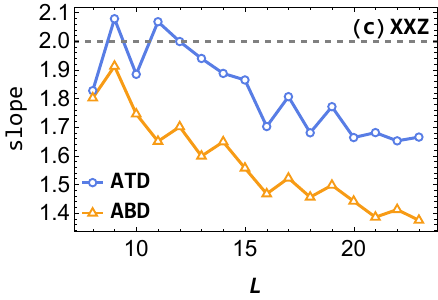}
  \caption{The average subsystem trace distance (left panel), Bures distance (middle panel), and fitted slope (right panel) in the sector with fixed momentum $K=1$ and magnetization $m = \frac{L}{2} - 2$ of the spin-1/2 XXZ chain. Here, $\Delta = \sqrt{2}$.
  The solid red lines are the function (\ref{universalfx}).}
  \label{FigureXXZ}
\end{figure}

In the appendix, we study the spin-1/2 XXZ chain with Hamiltonian
\begin{equation}
H = -\frac{1}{4} \sum_{j=1}^L \left( \sigma_j^x \sigma_{j+1}^x + \sigma_j^y \sigma_{j+1}^y + \Delta\,\sigma_j^z \sigma_{j+1}^z \right) - \frac{h_z}{2} \sum_{j=1}^L \sigma_j^z,
\end{equation}
subject to periodic boundary conditions \(\sigma_{L+1}^\alpha = \sigma_1^\alpha\) for \(\alpha = x,y,z\). The model is integrable via the Bethe ansatz \cite{Caux:2014uuq,Franchini:2016cxs}. In practice, we perform numerical block-diagonalization of the Hamiltonian for fixed momentum \(p = \frac{2\pi K}{L}\) and magnetization \(m = \frac{1}{2} \sum_{j=1}^L \langle \sigma_j^z \rangle\), following the procedure in \cite{Sandvik:2010lkj}. Imposing the magnetization constraint allows us to reach larger system sizes than those in \cite{Khasseh:2023kxw}. The GGE states of the spin-1/2 XXZ chain are determined not only by local conserved charges but also by quasilocal ones \cite{Ilievski:2015nca,Ilievski:2015tgk,Ilievski:2016fdy}. For simplicity, we compute the average subsystem distance between each pair of states within the sector with fixed \(K\) and \(m\). It is straightforward to see that the results are independent of the field \(h_z\). The results are shown in figure~\ref{FigureXXZ}. The data favor a slope smaller than 2, though the precise value remains to be further investigated.

\section{Universal slope 2?} \label{appSlo}

It was conjectured in \cite{Khasseh:2023kxw} that there exists a universal slope of 2 for the average subsystem trace distance \(\langle D_A \rangle\) in quadratic integrable systems. The panels (a), (e), and (f) of figure~\ref{FigureAD} are consistent with the conjecture for the average subsystem trace distance \(\langle D_A \rangle\) and scaled Bures distance \(\langle B_A \rangle/\sqrt{2}\), while panels (b) and (c) show slight deviations from the conjecture, and panels (d) and (h) exhibit more pronounced differences. Figure~\ref{FigureATDABTrandom} demonstrates that the conjecture does not apply to random pure Gaussian states, suggesting that the ordering of Gaussian states may play an essential role in the behavior of average subsystem distances. The results of the average subsystem trace distance in figure~\ref{FigureXXZ} for the spin-1/2 XXZ chain indicate a slope smaller than 2, while the average subsystem Bures distance seems not linear at all. Given these findings, the universality of the slope 2 remains an open question.

\begin{figure}[t]
  \centering
  \includegraphics[width=0.25\textwidth]{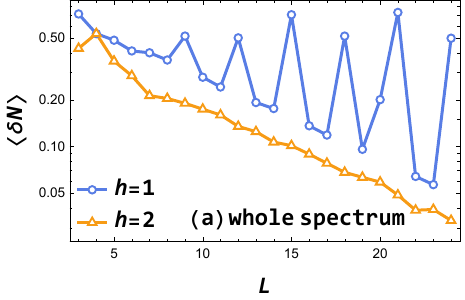} ~
  \includegraphics[width=0.25\textwidth]{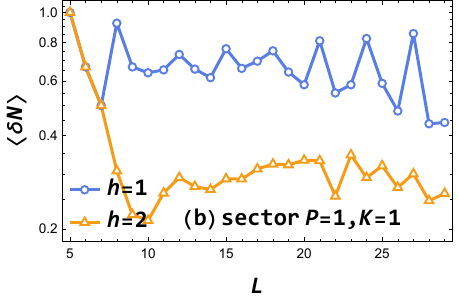}
  \caption{The average difference in the number of excited modes between neighboring states in the whole spectrum (left panel) and in the sector with fixed \( P=1, K=1 \) (right panel) of the Ising chain.}
  \label{FigureNumberDifference}
\end{figure}

If the universal slope of 2 is true for the Ising chain, it raises the question of how to understand it. One possible argument is that on average, the difference in the number of excited modes between neighboring states
\be
\langle \d N \rangle = \frac{1}{d-1} \sum_{i=1}^{d-1} | N_i - N_{i+1} |,
\ee
is 2. Note that $N_i$ is the number of modes in the state $|i\rag$. This is hypothesized to arise from the small interval expansion of the subsystem distance between quasiparticle excited states \cite{Zhang:2022tgu}, which could lead to the universal slope of 2. However, this argument does not hold. In figure~\ref{FigureNumberDifference}, we show the average mode number difference between neighboring states, which is much smaller than 2.


\begin{thebibliography}{10}

\bibitem{Caux:2010by}
J.-S. Caux and J.~Mossel, \textit{{Remarks on the notion of quantum
  integrability}}, \href{http://dx.doi.org/10.1088/1742-5468/2011/02/P02023}{J.
  Stat. Mech. (2011) P02023},
  [\href{https://arxiv.org/abs/1012.3587}{{\ttfamily arXiv:1012.3587}}].

\bibitem{Berry:1977kiw}
M.~V. {Berry} and M.~{Tabor}, \textit{{Level Clustering in the Regular
  Spectrum}}, \href{http://dx.doi.org/10.1098/rspa.1977.0140}{Proc. R. Soc.
  Lond. A. {\bfseries 356}, 375--394 (1977)}.

\bibitem{Bohigas:1984vrm}
O.~{Bohigas}, M.~J. {Giannoni} and C.~{Schmit}, \textit{{Characterization of
  Chaotic Quantum Spectra and Universality of Level Fluctuation Laws}},
  \href{http://dx.doi.org/10.1103/PhysRevLett.52.1}{Phys. Rev. Lett. {\bfseries
  52}, 1--4 (1984)}.

\bibitem{Finkel:2005yof}
F.~{Finkel} and A.~{Gonz{\'a}lez-L{\'o}pez}, \textit{{Global properties of the
  spectrum of the Haldane-Shastry spin chain}},
  \href{http://dx.doi.org/10.1103/PhysRevB.72.174411}{Phys. Rev. B {\bfseries
  72}, 174411 (2005)},
  [\href{https://arxiv.org/abs/cond-mat/0509032}{{\ttfamily
  arXiv:cond-mat/0509032}}].

\bibitem{Sieberer:2019xfn}
L.~M. {Sieberer}, T.~{Olsacher}, A.~{Elben}, M.~{Heyl}, P.~{Hauke}, F.~{Haake}
  and P.~{Zoller}, \textit{{Digital quantum simulation, Trotter errors, and
  quantum chaos of the kicked top}},
  \href{http://dx.doi.org/10.1038/s41534-019-0192-5}{npj Quantum Inf.
  {\bfseries 5}, 78 (2019)},
  [\href{https://arxiv.org/abs/1812.05876}{{\ttfamily arXiv:1812.05876}}].

\bibitem{Larkin:1969ifp}
A.~I. {Larkin} and Y.~N. {Ovchinnikov}, \textit{{Quasiclassical Method in the
  Theory of Superconductivity}}, \href{https://jetp.ras.ru/cgi-bin/e/index/e/28/6/p1200?a=list}{Sov. Phys. JETP {\bfseries 28}, 1200 (1969)}.





\bibitem{Maldacena:2015waa}
J.~Maldacena, S.~H. Shenker and D.~Stanford, \textit{{A bound on chaos}},
  \href{http://dx.doi.org/10.1007/JHEP08(2016)106}{JHEP {\bfseries 08} (2016)
  106}, [\href{https://arxiv.org/abs/1503.01409}{{\ttfamily
  arXiv:1503.01409}}].

\bibitem{Kukuljan:2017xag}
I.~Kukuljan, S.~Grozdanov and T.~Prosen, \textit{{Weak Quantum Chaos}},
  \href{http://dx.doi.org/10.1103/PhysRevB.96.060301}{Phys. Rev. B {\bfseries
  96}, 060301 (2017)}, [\href{https://arxiv.org/abs/1701.09147}{{\ttfamily
  arXiv:1701.09147}}].

\bibitem{Rozenbaum:2019nwn}
E.~B. Rozenbaum, L.~A. Bunimovich and V.~Galitski, \textit{{Early-Time
  Exponential Instabilities in Nonchaotic Quantum Systems}},
  \href{http://dx.doi.org/10.1103/PhysRevLett.125.014101}{Phys. Rev. Lett.
  {\bfseries 125}, 014101 (2020)},
  [\href{https://arxiv.org/abs/1902.05466}{{\ttfamily arXiv:1902.05466}}].

\bibitem{Vidmar:2017uux}
L.~Vidmar, L.~Hackl, E.~Bianchi and M.~Rigol, \textit{{Entanglement Entropy of
  Eigenstates of Quadratic Fermionic Hamiltonians}},
  \href{http://dx.doi.org/10.1103/PhysRevLett.119.020601}{Phys. Rev. Lett.
  {\bfseries 119}, 020601 (2017)},
  [\href{https://arxiv.org/abs/1703.02979}{{\ttfamily arXiv:1703.02979}}].

\bibitem{Vidmar:2017pak}
L.~Vidmar and M.~Rigol, \textit{{Entanglement Entropy of Eigenstates of Quantum
  Chaotic Hamiltonians}},
  \href{http://dx.doi.org/10.1103/PhysRevLett.119.220603}{Phys. Rev. Lett.
  {\bfseries 119}, 220603 (2017)},
  [\href{https://arxiv.org/abs/1708.08453}{{\ttfamily arXiv:1708.08453}}].

\bibitem{Vidmar:2018rqk}
L.~Vidmar, L.~Hackl, E.~Bianchi and M.~Rigol, \textit{{Volume Law and Quantum
  Criticality in the Entanglement Entropy of Excited Eigenstates of the Quantum
  Ising Model}}, \href{http://dx.doi.org/10.1103/PhysRevLett.121.220602}{Phys.
  Rev. Lett. {\bfseries 121}, 220602 (2018)},
  [\href{https://arxiv.org/abs/1808.08963}{{\ttfamily arXiv:1808.08963}}].

\bibitem{Hackl:2018tyl}
L.~Hackl, L.~Vidmar, M.~Rigol and E.~Bianchi, \textit{{Average eigenstate
  entanglement entropy of the XY chain in a transverse field and its
  universality for translationally invariant quadratic fermionic models}},
  \href{http://dx.doi.org/10.1103/PhysRevB.99.075123}{Phys. Rev. B {\bfseries
  99}, 075123 (2019)}, [\href{https://arxiv.org/abs/1812.08757}{{\ttfamily
  arXiv:1812.08757}}].

\bibitem{Jafarizadeh:2019xxc}
A.~Jafarizadeh and M.~Rajabpour, \textit{{Bipartite entanglement entropy of the
  excited states of free fermions and harmonic oscillators}},
  \href{http://dx.doi.org/10.1103/PhysRevB.100.165135}{Phys. Rev. B {\bfseries
  100}, 165135 (2019)}, [\href{https://arxiv.org/abs/1907.09806}{{\ttfamily
  arXiv:1907.09806}}].

\bibitem{LeBlond:2019eoe}
T.~LeBlond, K.~Mallayya, L.~Vidmar and M.~Rigol, \textit{{Entanglement and
  matrix elements of observables in interacting integrable systems}},
  \href{http://dx.doi.org/10.1103/PhysRevE.100.062134}{Phys. Rev. E {\bfseries
  100}, 062134 (2019)}, [\href{https://arxiv.org/abs/1909.09654}{{\ttfamily
  arXiv:1909.09654}}].

\bibitem{Kliczkowski:2023qmp}
M.~Kliczkowski, R.~{\'S}wi{\k{e}}tek, L.~Vidmar and M.~Rigol, \textit{{Average
  entanglement entropy of midspectrum eigenstates of quantum-chaotic
  interacting Hamiltonians}},
  \href{http://dx.doi.org/10.1103/PhysRevE.107.064119}{Phys. Rev. E {\bfseries
  107}, 064119 (2023)}, [\href{https://arxiv.org/abs/2303.13577}{{\ttfamily
  arXiv:2303.13577}}].

\bibitem{Page:1993df}
D.~N. Page, \textit{{Average entropy of a subsystem}},
  \href{http://dx.doi.org/10.1103/PhysRevLett.71.1291}{Phys. Rev. Lett.
  {\bfseries 71}, 1291--1294 (1993)},
  [\href{https://arxiv.org/abs/gr-qc/9305007}{{\ttfamily
  arXiv:gr-qc/9305007}}].

\bibitem{Foong:1994eja}
S.~K. Foong and S.~Kanno, \textit{{Proof of Page{\textquoteright}s conjecture
  on the average entropy of a subsystem}},
  \href{http://dx.doi.org/10.1103/PhysRevLett.72.1148}{Phys. Rev. Lett.
  {\bfseries 72}, 1148 (1994)}.

\bibitem{Sanchez-Ruiz:1995bhf}
J.~S{\'a}nchez-Ruiz, \textit{{Simple proof of Page{\textquoteright}s conjecture
  on the average entropy of a subsystem}},
  \href{http://dx.doi.org/10.1103/PhysRevE.52.5653}{Phys. Rev. E {\bfseries
  52}, 5653 (1995)}.

\bibitem{Sen:1996ph}
S.~Sen, \textit{{Average entropy of a subsystem}},
  \href{http://dx.doi.org/10.1103/PhysRevLett.77.1}{Phys. Rev. Lett. {\bfseries
  77}, 1--3 (1996)}, [\href{https://arxiv.org/abs/hep-th/9601132}{{\ttfamily
  arXiv:hep-th/9601132}}].

\bibitem{Liu:2017kfa}
C.~Liu, X.~Chen and L.~Balents, \textit{{Quantum Entanglement of the
  Sachdev-Ye-Kitaev Models}},
  \href{http://dx.doi.org/10.1103/PhysRevB.97.245126}{Phys. Rev. B {\bfseries
  97}, 245126 (2018)}, [\href{https://arxiv.org/abs/1709.06259}{{\ttfamily
  arXiv:1709.06259}}].

\bibitem{Zhang:2020kia}
P.~Zhang, C.~Liu and X.~Chen, \textit{{Subsystem R{\'e}nyi Entropy of Thermal
  Ensembles for SYK-like models}},
  \href{http://dx.doi.org/10.21468/SciPostPhys.8.6.094}{SciPost Phys.
  {\bfseries 8}, 094 (2020)},
  [\href{https://arxiv.org/abs/2003.09766}{{\ttfamily arXiv:2003.09766}}].

\bibitem{Lydzba:2020qfx}
P.~\L{}yd\.zba, M.~Rigol and L.~Vidmar, \textit{{Eigenstate Entanglement
  Entropy in Random Quadratic Hamiltonians}},
  \href{http://dx.doi.org/10.1103/PhysRevLett.125.180604}{Phys. Rev. Lett.
  {\bfseries 125}, 180604 (2020)},
  [\href{https://arxiv.org/abs/2006.11302}{{\ttfamily arXiv:2006.11302}}].

\bibitem{Lydzba:2021hml}
P.~\L{}yd\.zba, M.~Rigol and L.~Vidmar, \textit{{Entanglement in many-body
  eigenstates of quantum-chaotic quadratic Hamiltonians}},
  \href{http://dx.doi.org/10.1103/PhysRevB.103.104206}{Phys. Rev. B {\bfseries
  103}, 104206 (2021)}, [\href{https://arxiv.org/abs/2101.05309}{{\ttfamily
  arXiv:2101.05309}}].

\bibitem{Bianchi:2021lnp}
E.~Bianchi, L.~Hackl and M.~Kieburg, \textit{{Page curve for fermionic Gaussian
  states}}, \href{http://dx.doi.org/10.1103/PhysRevB.103.L241118}{Phys. Rev. B
  {\bfseries 103}, L241118 (2021)},
  [\href{https://arxiv.org/abs/2103.05416}{{\ttfamily arXiv:2103.05416}}].

\bibitem{Bianchi:2021aui}
E.~Bianchi, L.~Hackl, M.~Kieburg, M.~Rigol and L.~Vidmar, \textit{{Volume-Law
  Entanglement Entropy of Typical Pure Quantum States}},
  \href{http://dx.doi.org/10.1103/PRXQuantum.3.030201}{PRX Quantum {\bfseries
  3}, 030201 (2022)}, [\href{https://arxiv.org/abs/2112.06959}{{\ttfamily
  arXiv:2112.06959}}].

\bibitem{Khasseh:2023kxw}
R.~Khasseh, J.~Zhang, M.~Heyl and M.~A. Rajabpour, \textit{{Identifying Quantum
  Many-Body Integrability and Chaos Using Eigenstate Trace Distances}},
  \href{http://dx.doi.org/10.1103/PhysRevLett.131.216701}{Phys. Rev. Lett.
  {\bfseries 131}, 216701 (2023)},
  [\href{https://arxiv.org/abs/2301.13218}{{\ttfamily arXiv:2301.13218}}].

\bibitem{Nielsen:2010oan}
M.~A. Nielsen and I.~L. Chuang, \textit{{Quantum Computation and Quantum
  Information}}.
\newblock Cambridge University Press, Cambridge, UK, 10th anniversary~ed.,
  2010,
  \href{http://dx.doi.org/10.1017/CBO9780511976667}{10.1017/CBO9780511976667}.

\bibitem{Deutsch:1991msp}
J.~M. Deutsch, \textit{{Quantum statistical mechanics in a closed system}},
  \href{http://dx.doi.org/10.1103/PhysRevA.43.2046}{Phys. Rev. A {\bfseries
  43}, 2046--2049 (1991)}.

\bibitem{Srednicki:1994mfb}
M.~Srednicki, \textit{{Chaos and Quantum Thermalization}},
  \href{http://dx.doi.org/10.1103/PhysRevE.50.888}{Phys. Rev. E {\bfseries 50},
  888--901 (1994)}, [\href{https://arxiv.org/abs/cond-mat/9403051}{{\ttfamily
  arXiv:cond-mat/9403051}}].

\bibitem{Rigol:2007mja}
M.~Rigol, V.~Dunjko and M.~Olshanii, \textit{Thermalization and its mechanism
  for generic isolated quantum systems},
  \href{http://dx.doi.org/10.1038/nature06838}{Nature {\bfseries 452}, 854--858
  (2008)}, [\href{https://arxiv.org/abs/0708.1324}{{\ttfamily
  arXiv:0708.1324}}].

\bibitem{Dymarsky:2016ntg}
A.~Dymarsky, N.~Lashkari and H.~Liu, \textit{{Subsystem eigenstate
  thermalization hypothesis}},
  \href{http://dx.doi.org/10.1103/PhysRevE.97.012140}{Phys. Rev. E {\bfseries
  97}, 012140 (2018)}, [\href{https://arxiv.org/abs/1611.08764}{{\ttfamily
  arXiv:1611.08764}}].

\bibitem{Kudler-Flam:2021rpr}
J.~Kudler-Flam, \textit{{Relative Entropy of Random States and Black Holes}},
  \href{http://dx.doi.org/10.1103/PhysRevLett.126.171603}{Phys. Rev. Lett.
  {\bfseries 126}, 171603 (2021)},
  [\href{https://arxiv.org/abs/2102.05053}{{\ttfamily arXiv:2102.05053}}].

\bibitem{Kudler-Flam:2021alo}
J.~Kudler-Flam, V.~Narovlansky and S.~Ryu, \textit{{Distinguishing Random and
  Black Hole Microstates}},
  \href{http://dx.doi.org/10.1103/PRXQuantum.2.040340}{PRX Quantum {\bfseries
  2}, 040340 (2021)}, [\href{https://arxiv.org/abs/2108.00011}{{\ttfamily
  arXiv:2108.00011}}].

\bibitem{deMiranda:2022rze}
J.~T. de~Miranda and T.~Micklitz, \textit{{Subsystem trace-distances of two
  random states}}, \href{http://dx.doi.org/10.1088/1751-8121/acc770}{J. Phys. A
  {\bfseries 56}, 175301 (2023)},
  [\href{https://arxiv.org/abs/2210.03213}{{\ttfamily arXiv:2210.03213}}].

\bibitem{Bures:1969rqp}
D.~Bures, \textit{{An extension of Kakutani's theorem on infinite product
  measures to the tensor product of semifinite $\omega^*$-algebras}},
  \href{https://doi.org/10.2307/1995012}{Trans. Amer. Math.
  Soc. {\bfseries 135}, 199--212 (1969)}.

\bibitem{Jozsa:1994mnd}
R.~{Jozsa}, \textit{{Fidelity for Mixed Quantum States}},
  \href{http://dx.doi.org/10.1080/09500349414552171}{J. Mod. Opt. {\bfseries
  41}, 2315--2323 (1994)}.

\bibitem{Fagotti:2010yr}
M.~Fagotti and P.~Calabrese, \textit{{Entanglement entropy of two disjoint
  blocks in XY chains}},
  \href{http://dx.doi.org/10.1088/1742-5468/2010/04/P04016}{J. Stat. Mech.
  (2010) P04016}, [\href{https://arxiv.org/abs/1003.1110}{{\ttfamily
  arXiv:1003.1110}}].

\bibitem{Banchi:2013uht}
L.~{Banchi}, P.~{Giorda} and P.~{Zanardi}, \textit{{Quantum
  information-geometry of dissipative quantum phase transitions}},
  \href{http://dx.doi.org/10.1103/PhysRevE.89.022102}{Phys. Rev. E {\bfseries
  89}, 022102 (2014)}, [\href{https://arxiv.org/abs/1305.4527}{{\ttfamily
  arXiv:1305.4527}}].

\bibitem{Baldwin:2022cjb}
A.~J. Baldwin and J.~A. Jones, \textit{{Efficiently computing the Uhlmann
  fidelity for density matrices}},
  \href{http://dx.doi.org/10.1103/PhysRevA.107.012427}{Phys. Rev. A {\bfseries
  107}, 012427 (2023)}, [\href{https://arxiv.org/abs/2211.02623}{{\ttfamily
  arXiv:2211.02623}}].

\bibitem{Zhang:2022nuh}
J.~Zhang and M.~A. Rajabpour, \textit{{Trace distance between fermionic
  Gaussian states from a truncation method}},
  \href{http://dx.doi.org/10.1103/PhysRevA.108.022414}{Phys. Rev. A {\bfseries
  108}, 022414 (2023)}, [\href{https://arxiv.org/abs/2210.11865}{{\ttfamily
  arXiv:2210.11865}}].

\bibitem{Vidal:2002rm}
G.~Vidal, J.~I. Latorre, E.~Rico and A.~Kitaev, \textit{{Entanglement in
  Quantum Critical Phenomena}},
  \href{http://dx.doi.org/10.1103/PhysRevLett.90.227902}{Phys. Rev. Lett.
  {\bfseries 90}, 227902 (2003)},
  [\href{https://arxiv.org/abs/quant-ph/0211074}{{\ttfamily
  arXiv:quant-ph/0211074}}].

\bibitem{Latorre:2003kg}
J.~I. Latorre, E.~Rico and G.~Vidal, \textit{{Ground state entanglement in
  quantum spin chains}}, \href{http://dx.doi.org/10.26421/QIC4.1}{Quant. Inf.
  Comput. {\bfseries 4}, 48 (2004)},
  [\href{https://arxiv.org/abs/quant-ph/0304098}{{\ttfamily
  arXiv:quant-ph/0304098}}].

\bibitem{Grady:1982foy}
M.~{Grady}, \textit{{Infinite set of conserved charges in the Ising model}},
  \href{http://dx.doi.org/10.1103/PhysRevD.25.1103}{{Phys. Rev. D} {\bfseries
  25}, 1103--1113 (1982)}.

\bibitem{Prosen:1998uvt}
T.~{Prosen}, \textit{{Quantum invariants of motion in a generic many-body
  system}}, \href{http://dx.doi.org/10.1088/0305-4470/31/37/004}{{J. Phys. A:
  Math. Gen.} {\bfseries 31}, L645--L653 (1998)},
  [\href{https://arxiv.org/abs/cond-mat/9803358}{{\ttfamily
  arXiv:cond-mat/9803358}}].

\bibitem{Fagotti:2013jzu}
M.~Fagotti and F.~H. Essler, \textit{Reduced density matrix after a quantum
  quench}, \href{http://dx.doi.org/10.1103/PhysRevB.87.245107}{Phys. Rev. B
  {\bfseries 87}, 245107 (2013)},
  [\href{https://arxiv.org/abs/1302.6944}{{\ttfamily arXiv:1302.6944}}].

\bibitem{Lieb:1961fr}
E.~H. Lieb, T.~Schultz and D.~Mattis, \textit{{Two soluble models of an
  antiferromagnetic chain}},
  \href{http://dx.doi.org/10.1016/0003-4916(61)90115-4}{Annals Phys. {\bfseries
  16}, 407 (1961)}.

\bibitem{Katsura:1962hqz}
S.~Katsura, \textit{{Statistical mechanics of the anisotropic linear Heisenberg
  model}}, \href{http://dx.doi.org/10.1103/PhysRev.127.1508}{{Phys. Rev.}
  {\bfseries 127}, 1508 (1962)}.

\bibitem{Pfeuty:1970ayt}
P.~Pfeuty, \textit{{The one-dimensional Ising model with a transverse field}},
  \href{http://dx.doi.org/10.1016/0003-4916(70)90270-8}{Annals Phys. {\bfseries
  57}, 79 (1970)}.

\bibitem{Sachdev:1992fk}
S.~Sachdev and J.~Ye, \textit{{Gapless spin fluid ground state in a random,
  quantum Heisenberg magnet}},
  \href{http://dx.doi.org/10.1103/PhysRevLett.70.3339}{Phys. Rev. Lett.
  {\bfseries 70}, 3339 (1993)},
  [\href{https://arxiv.org/abs/cond-mat/9212030}{{\ttfamily
  arXiv:cond-mat/9212030}}].

\bibitem{Kitaev:2015qjp}
A.~Kitaev, ``A simple model of quantum holography.'' Talk presented at the
  {KITP} Program: Entanglement in Strongly-Correlated Quantum Matter, 2015.
\newblock
  [\href{http://online.kitp.ucsb.edu/online/entangled15/kitaev}{{Online}}].

\bibitem{Sachdev:2015efa}
S.~Sachdev, \textit{{Bekenstein-Hawking Entropy and Strange Metals}},
  \href{http://dx.doi.org/10.1103/PhysRevX.5.041025}{Phys. Rev. X {\bfseries
  5}, 041025 (2015)}, [\href{https://arxiv.org/abs/1506.05111}{{\ttfamily
  arXiv:1506.05111}}].

\bibitem{Polchinski:2016xgd}
J.~Polchinski and V.~Rosenhaus, \textit{{The Spectrum in the Sachdev-Ye-Kitaev
  Model}}, \href{http://dx.doi.org/10.1007/JHEP04(2016)001}{JHEP {\bfseries 04}
  (2016) 001}, [\href{https://arxiv.org/abs/1601.06768}{{\ttfamily
  arXiv:1601.06768}}].

\bibitem{Maldacena:2016hyu}
J.~Maldacena and D.~Stanford, \textit{{Remarks on the Sachdev-Ye-Kitaev
  model}}, \href{http://dx.doi.org/10.1103/PhysRevD.94.106002}{Phys. Rev. D
  {\bfseries 94}, 106002 (2016)},
  [\href{https://arxiv.org/abs/1604.07818}{{\ttfamily arXiv:1604.07818}}].

\bibitem{Vidmar:2016laa}
L.~Vidmar and M.~Rigol, \textit{Generalized gibbs ensemble in integrable
  lattice models}, \href{http://dx.doi.org/10.1088/1742-5468/2016/06/064007}{J.
  Stat. Mech. (2016) 064007},
  [\href{https://arxiv.org/abs/1604.03990}{{\ttfamily arXiv:1604.03990}}].

\bibitem{Cassidy:2011smq}
A.~C. {Cassidy}, C.~W. {Clark} and M.~{Rigol}, \textit{{Generalized
  Thermalization in an Integrable Lattice System}},
  \href{http://dx.doi.org/10.1103/PhysRevLett.106.140405}{Phys. Rev. Lett.
  {\bfseries 106}, 140405 (2011)},
  [\href{https://arxiv.org/abs/1008.4794}{{\ttfamily arXiv:1008.4794}}].

\bibitem{Wang:2022ufp}
Q.-Q. {Wang}, S.-J. {Tao}, W.-W. {Pan}, Z.~{Chen}, G.~{Chen}, K.~{Sun}, J.-S.
  {Xu}, X.-Y. {Xu}, Y.-J. {Han}, C.-F. {Li} and G.-C. {Guo},
  \textit{{Experimental verification of generalized eigenstate thermalization
  hypothesis in an integrable system}},
  \href{http://dx.doi.org/10.1038/s41377-022-00887-5}{Light. Sci. Appl.
  {\bfseries 11}, 194 (2022)}.

\bibitem{Caux:2014uuq}
J.-S. Caux, \textit{The Bethe Wavefunction}.
\newblock Cambridge University Press, 2014,
  \href{http://dx.doi.org/10.1017/CBO9781107053885}{10.1017/CBO9781107053885}.
  [Translated from M.~Gaudin,
  \textit{La Fonction d'Onde de Bethe}. Masson, 1983.]

\bibitem{Franchini:2016cxs}
F.~Franchini, \textit{{An introduction to integrable techniques for
  one-dimensional quantum systems}},
  \href{http://dx.doi.org/10.1007/978-3-319-48487-7}{Lect. Notes Phys.
  {\bfseries 940}, (2017)}, [\href{https://arxiv.org/abs/1609.02100}{{\ttfamily
  arXiv:1609.02100}}].

\bibitem{Sandvik:2010lkj}
A.~W. Sandvik, \textit{{Computational Studies of Quantum Spin Systems}},
  \href{http://dx.doi.org/10.1063/1.3518900}{AIP Conf. Proc. {\bfseries 1297},
  135 (2010)}, [\href{https://arxiv.org/abs/1101.3281}{{\ttfamily
  arXiv:1101.3281}}].

\bibitem{Ilievski:2015nca}
E.~Ilievski, M.~Medenjak and T.~Prosen, \textit{{Quasilocal Conserved Operators
  in the Isotropic Heisenberg Spin-1/2 Chain}},
  \href{http://dx.doi.org/10.1103/PhysRevLett.115.120601}{Phys. Rev. Lett.
  {\bfseries 115}, 120601 (2015)},
  [\href{https://arxiv.org/abs/1506.05049}{{\ttfamily arXiv:1506.05049}}].

\bibitem{Ilievski:2015tgk}
E.~{Ilievski}, J.~{De Nardis}, B.~{Wouters}, J.~S. {Caux}, F.~H.~L. {Essler}
  and T.~{Prosen}, \textit{{Complete Generalized Gibbs Ensembles in an
  Interacting Theory}},
  \href{http://dx.doi.org/10.1103/PhysRevLett.115.157201}{Phys. Rev. Lett.
  {\bfseries 115}, 157201 (2015)},
  [\href{https://arxiv.org/abs/1507.02993}{{\ttfamily arXiv:1507.02993}}].

\bibitem{Ilievski:2016fdy}
E.~Ilievski, M.~Medenjak, T.~Prosen and L.~Zadnik, \textit{{Quasilocal charges
  in integrable lattice systems}},
  \href{http://dx.doi.org/10.1088/1742-5468/2016/06/064008}{J. Stat. Mech.
  (2016) 064008}, [\href{https://arxiv.org/abs/1603.00440}{{\ttfamily
  arXiv:1603.00440}}].

\bibitem{Zhang:2022tgu}
J.~Zhang and M.~A. Rajabpour, \textit{{Subsystem distances between
  quasiparticle excited states}},
  \href{http://dx.doi.org/10.1007/JHEP07(2022)119}{JHEP {\bfseries 07} (2022)
  119}, [\href{https://arxiv.org/abs/2202.11448}{{\ttfamily
  arXiv:2202.11448}}].

\end{thebibliography}

\providecommand{\href}[2]{#2}\begingroup\raggedright\endgroup

\end{document}